%% file: Manuscript.tex
\documentclass[aps,prl,reprint,superscriptaddress, longbibliography,nofootinbib]{revtex4-2}
\usepackage[T1]{fontenc}
\usepackage[utf8]{inputenc}
\usepackage[english]{babel}   
\usepackage{lmodern}
\usepackage{amsmath,amssymb,amsthm,bm,bbm}
\usepackage{siunitx}
\usepackage{xspace}
\usepackage{graphicx}
\usepackage{hyperref}
\usepackage{braket}
\usepackage{physics}        
\usepackage{enumitem}       
\usepackage{times}
\usepackage{siunitx}  
\usepackage{listings}
\usepackage{xcolor} 
\usepackage{mathtools }
\usepackage{tabularx}
\usepackage{tabularray}
\usepackage{threeparttable}
\usepackage{orcidlink} 
\usepackage{tcolorbox}
\newcommand{\rem}[1]{{\color{black}#1}}
\newcommand{\remN}[1]{{\color{black}#1}}
\begin{document}
\title{Geometry-Controlled Freezing and Revival of Bell Nonlocality \\through Environmental Memory}

\author{Mohamed Hatifi\,\orcidlink{0009-0005-3368-2751}}
\email{hatifi@fresnel.fr}
\affiliation{Aix Marseille Univ, CNRS, Centrale M\'editerran\'ee, Institut Fresnel, Marseille, France}

\date{\today}
\begin{abstract}
We show that the distance between two qubits coupled to a structured reservoir acts as a single geometric control that can store, revive, or suppress Bell nonlocality. In a mirror-terminated guide, quantum correlations lost to the bath return at discrete recurrence times, turning a product state into a Bell-violating one without any entangling drive (only local basis rotations/readout). In the continuum limit, we derive closed-form criteria for the emergence of nonlocality from backflow, and introduce a Bell-based analogue of the BLP measure to quantify this effect. We also show how subwavelength displacements away from a decoherence-free node quadratically reduce the lifetime of a dark state or bright state, enabling highly sensitive interferometric detection. All results rely on analytically solvable models and are compatible with current superconducting and nanophotonic platforms, offering a practical route to passive, geometry-controlled non-Markovian devices.
\end{abstract}

\maketitle

\paragraph{Introduction.---}

Quantum nonlocality, witnessed by Bell-inequality violations, is a defining feature of quantum mechanics and underpins quantum communication, secure randomness, and device-independent protocols \cite{collins2002,kuzmich2003,Tanzilli_photonic2005,pomarico2011pra,bernien2013,giustina2013,hensen2015,hatifi2022}. In realistic platforms, coupling to the electromagnetic environment degrades correlations; under local completely positive divisible  \rem{(CP-divisible, i.e., Markovian)} noise, the CHSH  \rem{(Clauser-Horne-Shimony-Holt)} parameter typically decreases and is not restored by local operations alone \cite{barrett2002,griffiths2020,nadlinger2022}. Recent progress in superconducting circuits, trapped ions, and nanophotonics has entered regimes where the bath retains memory and information backflow can partially recover lost coherence \cite{sillanpaa2007,devoret2013,cirac1995,harty2014,julsgaard2004,lvovsky2009,chruscinski2011,eichler2012,shankar2013,ballance2015,malekakhlagh2016,sampaio2017}.
Recurrences have been observed numerically in discrete-mode environments \cite{buzek1999,hatifi2022,hansen2023}, yet whether Bell nonlocality itself can revive after environmental decoherence--and how such revivals depend on geometric and spectral parameters--has remained open \cite{lorenzo2013,sampaio2017}. Answering this question would identify regimes where entanglement is regenerated through tailored system--environment interference, paving the way for device-independent protocols that harness non-Markovian effects. 
\\
Here we show, analytically and numerically, that emitter separation alone can store, revive, or quench Bell nonlocality in systems coupled to regularly spaced discrete baths (e.g., finite-length waveguides or cavity arrays). A mapping to a mirror-terminated waveguide explains periodic Bell revivals at integer multiples of the round-trip time. Taking the continuum limit, an effective four-mode embedding exactly reproduces a Lorentzian memory kernel and yields compact criteria for Bell violation \rem{ that link distance $d$ and bandwidth $\lambda$ to the timing and visibility of CHSH revivals}. \rem{Within the explored regime, we show that increases of the CHSH parameter coincide with information backflow, establishing a direct quantitative connection between Bell-inequality revival and non-Markovian backflow.} To quantify the effect, we introduce a Bell-based analogue of the Breuer-Laine-Piilo \rem{(BLP)} measure and show that its peaks \rem{align} with CHSH revival and mutual-information backflow \cite{breuer2009,chruscinski2011,eichler2012,shankar2013}\rem{, providing a Bell-specific, experimentally accessible witness of memory effects}. \rem{We also show that geometry enables a passive interferometric application: subwavelength displacements from a symmetry-protected node produce a quadratic change in dark/bright decay rates, giving simple design rules. These define operating windows in the $(d,\lambda)$ \emph{plane} that fix sensitivity and dynamic range and realize a quantum nondemolition strain sensor.} \remN{In this regime, the protected lifetime $T_{\mathrm{df}}\!\equiv\!\Gamma_{\mathrm{df}}^{-1}$ scales as $T_{\mathrm{df}}\propto(\delta d)^{-2}$ for $|k_0\delta d|\!\ll\!1$. Here $k_0=\omega_0/c$ and $\delta d$ denotes the small displacement from a coupling node (see below for definitions; cf. Eq.~\eqref{eq:Tdf_exact}). This yields a direct, calibration-free map from displacement to time.} 
Throughout, we analyze the single-excitation regime relevant for weak driving. The theory is analytically tractable and compatible with current superconducting and nanophotonic capabilities, outlining a practical, geometry-controlled route to non-Markovian devices without entangling drives\rem{--requiring only static geometry and local basis rotations/readout--}\rem{and with parameter sets suitable for immediate experimental tests}. \remN{Unlike prior studies of non-Markovian revivals in discrete baths \cite{buzek1999,hansen2023,lorenzo2013,sampaio2017}, we \emph{certify} Bell nonlocality, derive \emph{closed-form} criteria linking emitter separation and reservoir bandwidth to the timing and visibility of CHSH revivals, and demonstrate a passive, geometry-defined device based on the same interference.}

\paragraph{Physical model.---}
We consider two identical qubits (two-level systems, TLSs) at positions $x_1=-d$ and $x_2=+d$ interacting with a \emph{finite, discrete set} of bosonic modes that represent the electromagnetic environment. This is equivalent to two qubits in a mirror-terminated one-dimensional waveguide of length $L$, whose boundary conditions quantize the field into a uniformly spaced ladder of standing-wave modes with $k_n=n\pi/L$, with group velocity $v$ and spacing $\delta\omega=\pi v/L$. We set $\hbar=1$ and work within the dipole and rotating-wave approximations. The Hamiltonian reads
\begin{align}
H &= \omega_0 \big( \sigma_1^+ \sigma_1^- + \sigma_2^+ \sigma_2^- \big)
  + J \big( \sigma_1^+ \sigma_2^- + \sigma_1^- \sigma_2^+ \big)
  + \sum_k \omega_k\, b_k^\dagger b_k \nonumber\\
  &\quad + \sum_k \Big[
      g_k^{(1)}\, \sigma_1^- b_k^\dagger
    + g_k^{(2)}\, \sigma_2^- b_k^\dagger
    + \mathrm{h.c.}
  \Big],
\label{eq:H_full}
\end{align}
where $\omega_0$ and $\omega_k$ are the qubit and mode frequencies, $\sigma_j^\pm$ are ladder operators for TLS $j=1,2$, and $b_k^\dagger$ creates a bath excitation. The exchange $J$ accounts for coherent dipole-dipole interaction \cite{kaufman2015}. The position dependence enters through the couplings $g_k^{(j)}$, which we take as $g_k^{(j)} = g_k e^{\mathrm{i}k x_j}$  \cite{buzek1999,shen2007,zhou2008,hatifi2022,hansen2023}.  Since the qubits predominantly interact with modes near their resonance $\omega_0$, to proceed analytically, we approximate $k \simeq k_0 = \omega_0/v$ in the phase factors only, yielding
\begin{equation}
g_k^{(1)} \simeq g_k\, e^{+i k_0 d}, \qquad
g_k^{(2)} \simeq g_k\, e^{-i k_0 d},
\label{eq:phases_k0}
\end{equation}
while keeping the exact dispersion $\omega_k$ elsewhere. The resulting approximation incurs an error $\mathcal{O}[(\Delta\omega\, d / v)^2]$, which remains below $10^{-6}$ for typical parameters (\(\Delta\omega/\omega_0 \sim 10^{-3},\; d/\lambda_0 \lesssim 1\)). This well-controlled substitution, standard in treatments of collective emission~\cite{lehmberg1970, scully1997}, preserves all interference effects and enables closed-form analytic solutions in both the discrete and continuous regimes.

\paragraph*{Symmetric-antisymmetric basis and interference.---}
Crucially, the total number of excitations,
\[
N_{\mathrm{ex}} = \sum_{j=1}^2 \sigma_j^+ \sigma_j^- + \sum_k b_k^\dagger b_k,
\]
is conserved by Eq.~\eqref{eq:H_full}, allowing the dynamics to be exactly solved within the single-excitation sector. 
\\
We express the total state as
\[
|\Psi(t)\rangle = \alpha_1(t)|eg,\{0\}\rangle + \alpha_2(t)|ge,\{0\}\rangle + \sum_k \beta_k(t) |gg\rangle \otimes |1_k\rangle,
\]
where \( |e\rangle_j \) (resp.  \( |g\rangle_j \)) denotes the excited (resp. ground) state of the TLS \( j \), and \( |1_k\rangle \) a single excitation in bath mode \( k \).  We also define collective amplitudes $s(t) = [\alpha_1(t) + \alpha_2(t)]/\sqrt{2}$ and $a(t) = [\alpha_1(t) - \alpha_2(t)]/\sqrt{2}$, associated with the symmetric and antisymmetric states \(|S\rangle = (|eg\rangle + |ge\rangle)/\sqrt{2}\) and \(|A\rangle = (|eg\rangle - |ge\rangle)/\sqrt{2}\), which diagonalize the coherent exchange term. Using Eq.~\eqref{eq:phases_k0}, the system-bath interaction becomes
\begin{equation}
H_{\mathrm{SB}} =
\sum_k \sqrt{2} g_k \left[ \cos(k_0 d)\, \sigma_S^+ + i \sin(k_0 d)\, \sigma_A^+ \right] b_k + \text{H.c.},
\label{eq:bright_dark_coupling}
\end{equation}
with collective operators $\sigma_{S,A}^+ = (\sigma_1^+ \pm \sigma_2^+)/\sqrt{2}$. This expression reveals that the two decay channels couple independently to the reservoir, with amplitudes set by the distance through $\cos(k_0 d)$ and $\sin(k_0 d)$. For specific separations $d = n\lambda_0/2$ or $(2n{+}1)\lambda_0/4$, one of the two states becomes completely decoupled from the bath by interferences, forming a geometry-protected dark state. Such dark states act as coherent population traps that can protect entanglement against decay and enable distance-based reservoir engineering of Bell nonlocality. Through this interference mechanism, the separation $d$ becomes a single, coherent control: varying $d$ reweights the bright and dark channels, thereby freezing, reviving, or quenching Bell nonlocality.
\begin{figure}[t]
\centering
\includegraphics[width=.5\textwidth]{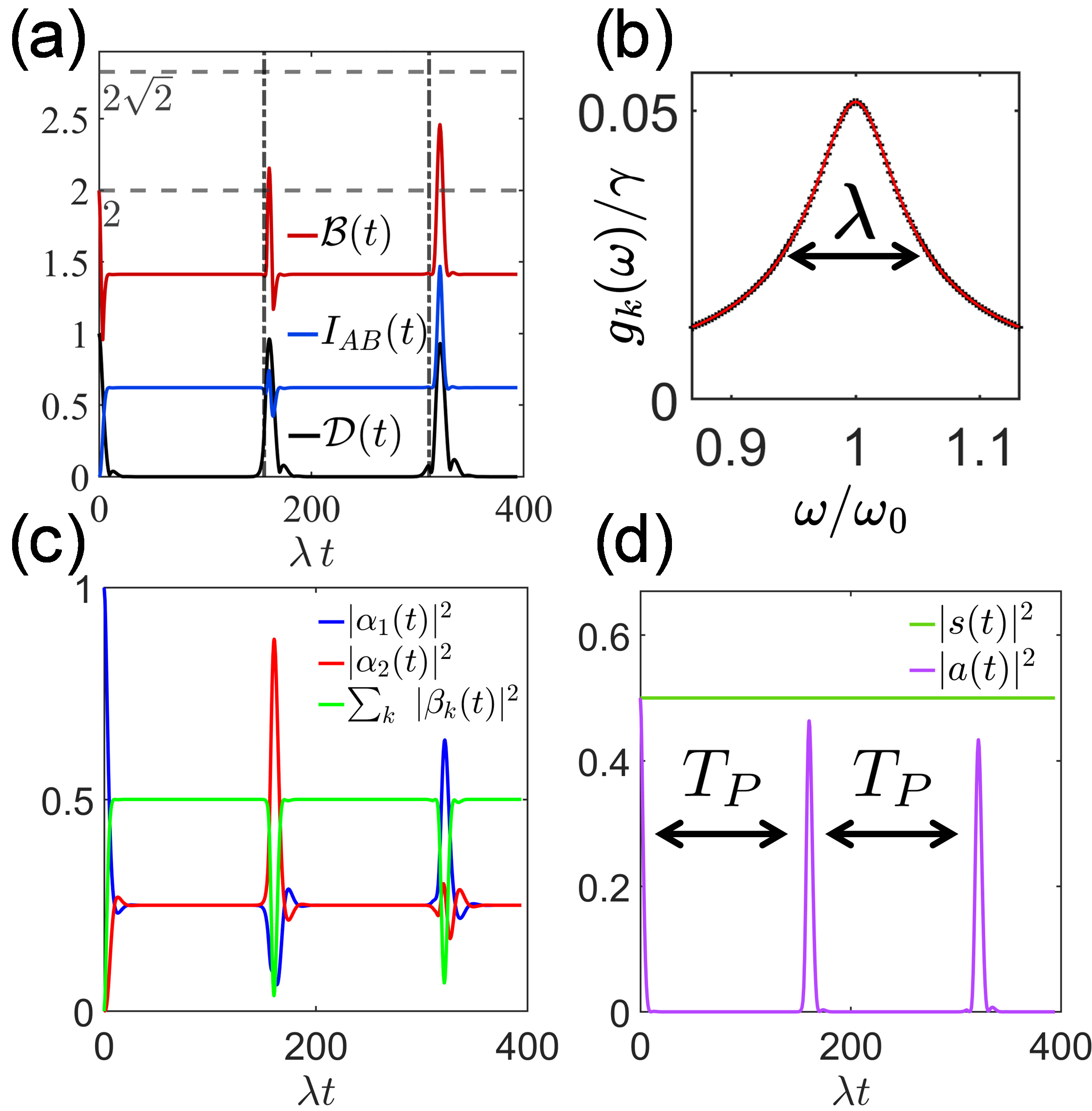}
\caption{%
\textbf{Geometry-controlled revival of Bell nonlocality in a discrete-mode bath.}
(a)~Time traces of the CHSH parameter $\mathcal{B}(t)$ (red), mutual information $I_{AB}(t)$ (blue), and trace distance $\mathcal{D}(t)$ (black) for two qubits coupled to a finite discrete bath. All peak at the Poincar\'e times $t=nT_P$ (vertical dashed), indicating information backflow and restoring Bell violation; grey lines mark $\mathcal{B}=2$ and $2\sqrt{2}$. The integrated backflow measures are $\mathcal{N}=2.41$ and $\mathcal{N}_B=2.55$. 
(b)~Discretised Lorentzian coupling spectrum $g_k(\omega)$ (crosses) matching the target envelope (solid line) with linewidth~$\lambda$. 
(c) Qubit populations $|\alpha_{1,2}(t)|^2$ (blue/red) and total bath population $\sum_k|\beta_k(t)|^2$ (green) show partial excitation return each period. (d) Collective states: the antisymmetric (bright) amplitude $|a(t)|^2$ (purple) revives at $nT_P$, while the symmetric (dark) $|s(t)|^2$ (green) remains constant. Initial state $|eg\rangle$ (no initial entanglement). Parameters: $d=\lambda_0/4$, $J=-10^{-3}\omega_0$, $N_m=100$, $\gamma=0.05\,\omega_0$, $\lambda=0.066\,\omega_0$.
}
\label{fig:revival_panel}
\end{figure}
\paragraph{Spectral structure and non-Markovianity.---}
We model the reservoir as a bosonic bath with a Lorentzian spectral density centred in $\omega_0$,
\begin{equation}
\mathcal{J}_L(\omega) = \frac{\gamma \lambda^2}{\left(\omega-\omega_0\right)^2 + \lambda^2},
\label{eq:lorentzian}
\end{equation}
where $\gamma$ sets the on-resonance system-bath coupling strength and $\lambda$ is the spectral linewidth. This form arises naturally in cavity QED \cite{ruggenthaler2014}, slow-light photonics \cite{baba2008}, and engineered microwave environments, and describes a reservoir with finite memory time $\tau_M \sim \lambda^{-1}$. Non-Markovian signatures appear when the bath correlation time $\tau_M \sim 1/\lambda$ becomes comparable to the system’s relaxation time $1/\gamma$, allowing coherent backaction that can temporarily restore lost correlations.

\paragraph*{Performance metrics.---}
To disentangle mere population echoes from genuine memory effects, we monitor three time-resolved quantities, all computable from the reduced two-qubit density matrix $\rho_{AB}(t)$.
\begin{figure}[t]
\centering
\includegraphics[width=\columnwidth]{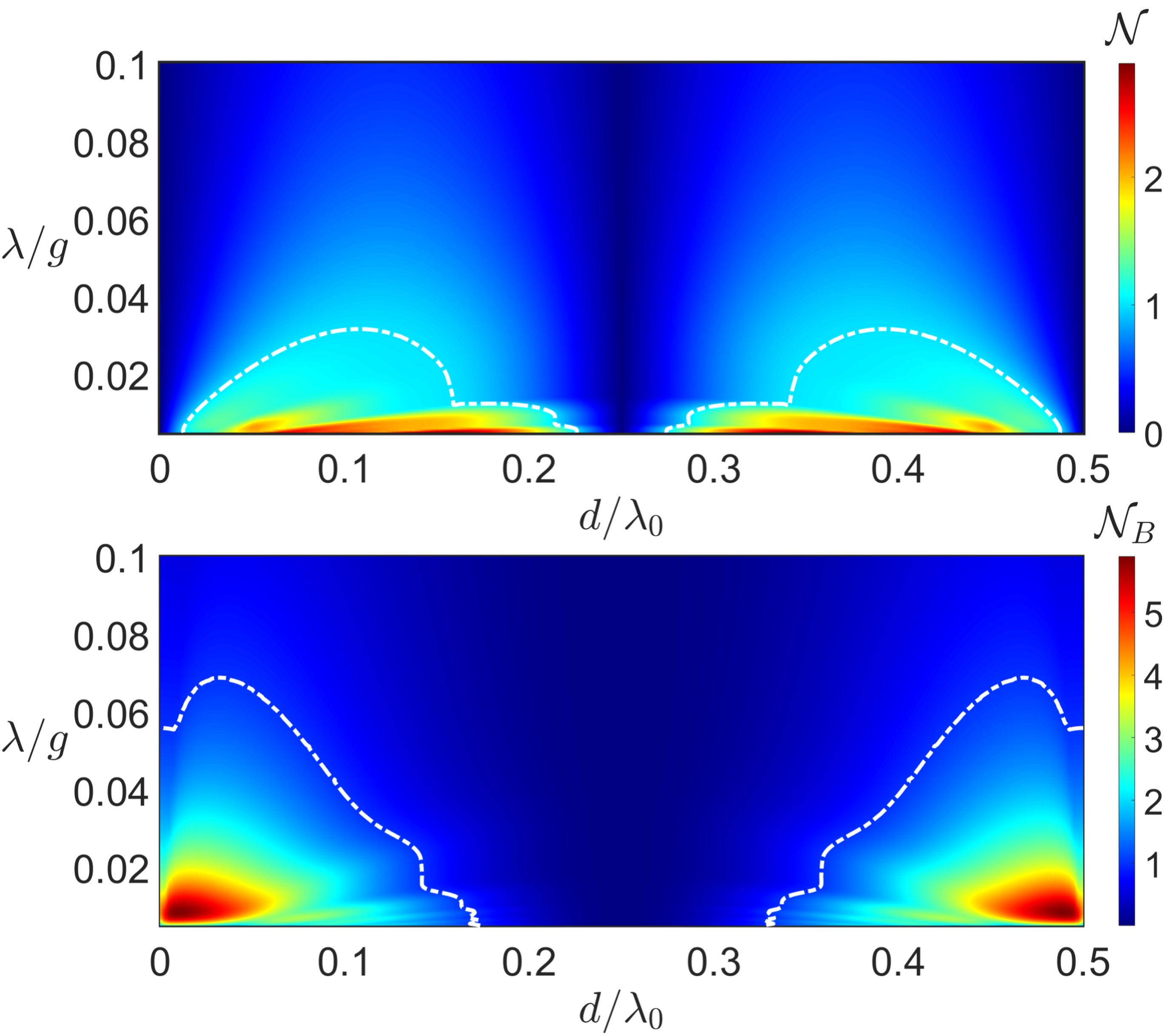}
\caption{
\textbf{Geometry-bandwidth map in the continuum.}
Maps of (a) BLP non-Markovianity $\mathcal{N}(d,\lambda)$ and (b) Bell backflow $\mathcal{N}_B(d,\lambda)$ obtained from the exact four-mode dynamics~\eqref{eq:cont_odes} for the symmetric Bell state $|S\rangle=(|eg\rangle+|ge\rangle)/\sqrt{2}$. Only one half-period is shown in $d \in [0,\lambda_0/2]$, as the dynamics are periodic in $d$ with period $\lambda_0$. The white dashed contour $\mathcal{N}=1$ marks the onset of coherent memory. The close overlap of the lobes in (a,b) indicates that Bell revivals occur in the same parameter regions where information backflow is present. Parameters: $\omega_0/2\pi=5\,\mathrm{GHz}$, $g=0.05\,\omega_0$, $J=-0.005\,\omega_0$.
}

\label{fig:3}
\end{figure}
\emph{(i) Trace-distance backflow.}  
For the orthogonal pair $\rho_{1,2}(0)=\{|eg\rangle\!\langle eg|,\,|ge\rangle\!\langle ge|\}$, which is optimal for detecting trace-distance backflow in the single-excitation manifold, the trace distance $D(t)=\tfrac12\lVert\rho_1(t)-\rho_2(t)\rVert_1$ admits a closed-form expression  \(D(t) = 2|a(t)\,s(t)|\). 

This quantity measures optimal state distinguishability, and its positive time-derivative defines the Breuer-Laine-Piilo (BLP) non-Markovianity functional \cite{breuer2009}
\[
\mathcal N
     =\!\int_{\dot D>0}\!\!\dot D(t)\,dt ,
\]
which vanishes for every CP-divisible (Markovian) channel.
\\
\emph{(ii) Bell backflow.}  
We introduce an analogous witness for the resurgence of nonlocal correlations,
\[
\mathcal{N}_{\mathrm{B}}
      = \int_{\dot{\mathcal{B}}>0} \dot{\mathcal{B}}(t)\,dt ,
\]
where \(\mathcal{B}(t)\) is the \emph{maximal} CHSH value achievable for \(\rho_{AB}(t)\), evaluated via the Horodecki criterion \cite{horodecki1995,simon2000,kaufman2015}. \\
In our simulations, increases of \(\mathcal{B}(t)\) occur only when the dynamics are non-CP-divisible, so a finite \(\mathcal{N}_{\mathrm{B}}\)  signals Bell backflow, i.e., restoration of nonlocality driven by the intrinsic system--bath dynamics, rather than mere passive preservation.

\emph{(iii) Quantum mutual information.}  
The total (classical + quantum) correlations are quantified by the quantum mutual information  
\(I_{AB}(t)=S\!\bigl(\rho_A(t)\bigr)+S\!\bigl(\rho_B(t)\bigr)-S\!\bigl(\rho_{AB}(t)\bigr)\),  
where \(S(\rho)\equiv-\operatorname{Tr}[\rho\log_2\rho]\) is the von Neumann entropy \cite{mazzola2010}.  
Peaks of \(I_{AB}(t)\) pinpoint the instants when information that has leaked to the environment flows back into the two-qubit register and is stored there, thereby corroborating the backflow witnesses above.

Figure~\ref{fig:revival_panel} demonstrates how geometry alone can steer an initially separable state into--and back out of--Bell nonlocality via coherent memory effects.  
The qubits are prepared in the product state \(|eg\rangle\), which decomposes into equal symmetric and antisymmetric amplitudes \((s,a) = (1/\sqrt{2},1/\sqrt{2})\), so no Bell violation is present at \(t = 0\).  
The reservoir is represented by \(N_m\) modes sampled from the Lorentzian spectral density of Eq.~\eqref{eq:lorentzian} (panel \textbf{b}).  
This discrete bath is exactly equivalent to a mirror-terminated one-dimensional waveguide of length \(L\), whose boundary conditions quantise the field to \(k_m = m\pi/L\) and impose a uniform frequency spacing \(\delta\omega = \pi v/L\) centred on \(\omega_0\).
In the single-excitation sector the system wavefunction becomes a finite Fourier series over the discrete bath modes.  
Because each mode acquires a phase that is an integer multiple of the spacing \(\delta\omega\), all phases re-align after the Poincar\'e time \(T_P \equiv 2L/v=2\pi/\delta\omega\) -- the single photon round-trip between the mirrors.  
When the reservoir linewidth satisfies the underdamped condition \(\lambda \lesssim \pi/T_P\), this global rephasing restores qubit-bath coherence and produces Bell revivals at every integer multiple of \(T_P\).  
At those instants (vertical dashed lines) energy, information, and nonlocal correlations flow back into the register: the excitation is partially reabsorbed by the qubits (panel \textbf{c}), and the three witnesses -- trace distance \(\mathcal{D}(t)\), mutual information \(I_{AB}(t)\), and CHSH parameter \(\mathcal{B}(t)\) -- peak simultaneously (panel \textbf{a}), with \(\mathcal{B}(t)\) exceeding the classical bound.
The emitters are separated by \(d=\lambda_0/4\), so that \(k_0 d = \pi/2\). Under this purely geometric condition, the \emph{symmetric} state decouples from the reservoir--turning the nominally bright channel into a dark, decoherence-free subspace--while the \emph{antisymmetric} state remains weakly coupled and undergoes the full non-Markovian round-trip.  
(Panel~\textbf{d}) confirms this behavior: the antisymmetric amplitude revives at each Poincar\'e time \(nT_P\), while the symmetric component stays frozen at its initial value. \rem{At nodal separations---$\sin(k_0 d)=0$ for $|A\rangle$ and $\cos(k_0 d)=0$ for $|S\rangle$---the corresponding Bell state decouples from the bath (see Eq.~\eqref{eq:bright_dark_coupling}). Thus, if the system is initialized in that state, its amplitude evolves trivially: the population---and hence the associated CHSH value---remains constant in time. Geometry therefore \emph{freezes} nonlocality without drive or feedback. Detuning the distance $d$ slightly away from a node switches the behavior: freezing at the node, revival at the Poincar\'e times $nT_P$, and quench when the bright channel dominates.} This distance-controlled routing of excitations closes a coherent, without entangling drive information loop: a separable state evolves into a Bell-violating one and returns to classicality, mediated entirely by reservoir memory structured through geometry.  
Decoherence, in this setting, is no longer synonymous with irreversibility--the bath acts as a quantum memory that temporarily stores and subsequently returns nonlocal correlations \cite{barrett2002}. 
Emitter separation thus becomes a passive, reversible, and geometry-tunable control for Bell nonlocality.
\paragraph{Continuum limit: exact four-mode pseudomode theory---} The discrete revivals discussed so far arise from the mode quantization imposed by finite mirror separation. 

To analyze the continuum regime,  we let the mirror separation tend to infinity, \(L\!\to\!\infty\), so the waveguide becomes an open continuum.  
The sum in Eq.~\eqref{eq:bright_dark_coupling} converts to a spectral integral,
\[
\sum_k |g_k|^{2}(\cdots)\;\longrightarrow\;
\int_{0}^{\infty}\!d\omega\,\mathcal{J}_{L}(\omega)\,(\cdots),
\]
with \(\mathcal{J}_{L}(\omega)\)  the Lorentzian density of states Eq.~\eqref{eq:lorentzian}.  
A Fourier transform yields an exponentially decaying memory kernel
\(K(t) = g^{2} e^{-\tilde{\lambda} t}\Theta(t)\),
where \(g = \sqrt{\gamma\lambda/2}\) and \(\tilde{\lambda} = \lambda - i\omega_{0}\).  
The reservoir therefore retains information for the finite correlation time \(\tau_{M} = \lambda^{-1}\), preserving possible non-Markovian dynamics even though strict Poincar\'e recurrences disappear as \(T_{P}=2L/v\to\infty\).
\paragraph*{Auxiliary amplitudes and local embedding.}
The exponential memory kernel \(K(t)\) permits an exact time-local reformulation. We define two pseudomode amplitudes that encode the reservoir’s memory,
\[
\beta(t)=ig\!\int_{0}^{t} e^{-\tilde{\lambda}(t-\tau)}\,s(\tau)\,d\tau,
\quad
\alpha(t)=ig\!\int_{0}^{t} e^{-\tilde{\lambda}(t-\tau)}\,a(\tau)\,d\tau.
\]
Together with the qubit amplitudes \(s(t)\) and \(a(t)\) they obey the exact linear set
\begin{align}
\dot s &= -i(\omega_0+J)\,s
        +2ig\cos^{2}(k_0 d)\,\beta
        -g\sin(2k_0 d)\,\alpha, \notag\\
\dot a &= -i(\omega_0-J)\,a
        +2ig\sin^{2}(k_0 d)\,\alpha
        +g\sin(2k_0 d)\,\beta, \notag\\
\dot\beta &= ig\,s-\tilde{\lambda}\,\beta,
\qquad
\dot\alpha = ig\,a-\tilde{\lambda}\,\alpha.
\label{eq:cont_odes}
\end{align}
Equations \eqref{eq:cont_odes} are \emph{Markovian in the four-mode embedding}: with the two memory pseudomodes explicit, the evolution is first-order and memoryless. Tracing them out restores the bath correlation time \(\tau_M=1/\lambda\), so the reduced qubit dynamics are non-Markovian for any finite \(\lambda\). This compact four-mode embedding is therefore exact for arbitrary separation \(2d\), linewidth \(\lambda\), exchange $J$, and coupling \(g\).
\paragraph*{Geometry-bandwidth map.---}
Figure~\ref{fig:3} charts the two backflow measures across the continuum parameter space, using the four-mode dynamics of Eq.~\eqref{eq:cont_odes}.  
The system is prepared in the symmetric Bell state \(|S\rangle=(|eg\rangle+|ge\rangle)/\sqrt{2}\), i.e.\ \(s(0)=1\) and \(a(0)=0\).  
Both the BLP non-Markovianity \(\mathcal{N}\) (panel~\textbf{a}) and the Bell backflow integral \(\mathcal{N}_{\mathrm{B}}\) (panel~\textbf{b}) display lobe-shaped regions of positive value that open at small reservoir linewidth \(\lambda\) and depend sensitively on the emitter separation \(2d\).  
These lobes mark parameter domains where coherent memory exchange between the qubits and the structured bath is sufficiently strong  to rebuild lost correlations.  
Only a half-period in distance, \(0\le d\le\lambda_{0}/2\), is shown for clarity; all observables are periodic in $d$ with period \(\lambda_{0}\).  
The white dashed contour traces \(\mathcal{N}=1\) and marks the onset of coherent memory effects: in both panels, the lobe boundaries coincide, confirming that revivals of Bell nonlocality occur in regions that support genuine information backflow. \rem{ Geometry-controlled storage and revival enable (i) drive-free preservation of nonlocality for device-independent tests and (ii) geometry-set revival times for synchronized Bell sampling. The criteria in Fig.~\ref{fig:3} translate directly to superconducting waveguides and nanophotonic devices, and the same mechanism extends to other structured continua (e.g., trapped-ion phonon modes and quantum dots in photonic crystals).}

\paragraph*{Exact diagonalization of the four-mode evolution matrix.---} In the continuum regime, the single-excitation dynamics reduce to a linear first-order system for the vector \( \bm X(t) = [s(t), a(t), \beta(t), \alpha(t)]^{\mathsf T} \), governed by
\begin{equation}
\dot{\bm X}(t) = M(d)\, \bm X(t),
\end{equation}
with the distance-dependent evolution matrix
\begin{equation}\small
M(d) = \begin{pmatrix}
-i(\omega_0+J) & 0 & 2ig\cos^2(k_0 d) & -g\sin(2k_0 d) \\
0 & -i(\omega_0-J) & g\sin(2k_0 d) & 2ig\sin^2(k_0 d) \\
ig & 0 & -\tilde\lambda & 0 \\
0 & ig & 0 & -\tilde\lambda
\end{pmatrix},
\label{eq:M_full}
\end{equation}
where \( \tilde\lambda = \lambda - i\omega_0 \). The matrix \( M(d) \) is diagonalizable for all real-valued\( (g, \lambda, J) \), and admits a similarity decomposition
\begin{equation}
M(d) = P(d)\,\Lambda(d)\,P^{-1}(d),
\quad \Lambda(d) = \mathrm{diag}(\mu_1, \mu_2, \mu_3, \mu_4),
\end{equation}
with eigenvectors \( \bm e_i(d) \) forming the columns of \( P(d) \), which map the physical amplitudes into the normal-mode basis, and eigenvalues \( \mu_i \in \mathbb{C} \) determining the full temporal behavior. In this basis, the evolution becomes
\begin{equation}
\bm X(t) = P(d)\, e^{\Lambda(d)t}\, P^{-1}(d)\, \bm X(0),
\end{equation}
which yields closed-form expressions for all physical observables.

In particular, the decoherence-protected lifetime is set by the slowest decay rate, \(T_{\mathrm{df}}^{-1}=-\mathrm{Re}\,\min_{i}\{\mu_{i}\}\), a quantity that directly sets the sensitivity of the passive, geometry-based strain sensor discussed below.
\begin{figure}[t]
\centering
\includegraphics[width=\columnwidth]{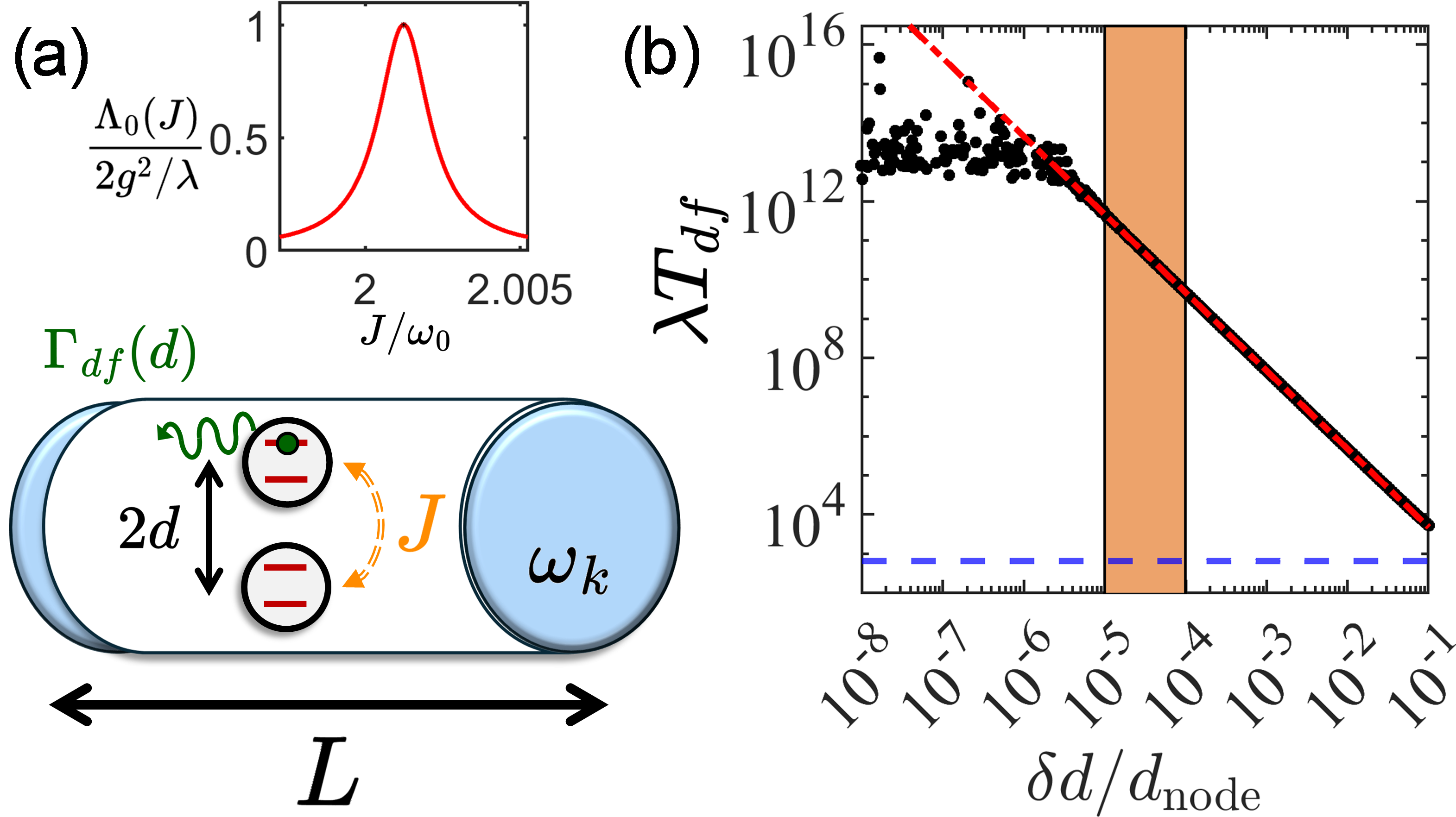}
\caption{%
\textbf{Strain sensing via geometry-protected Bell states.}
(a)~Concept: two identical qubits embedded in a mirror-terminated waveguide of length \(L\) are separated by \(2d\).
Tuning \(d\) places, respectively, the symmetric Bell state \(|S\rangle = (|eg\rangle + |ge\rangle)/\sqrt2\) or the antisymmetric state \(|A\rangle = (|eg\rangle - |ge\rangle)/\sqrt2\) at a collective-coupling node, rendering it completely dark.
Inset: normalized dark-state decay prefactor \(\Lambda_0(J)/(2g^2/\lambda)\) from Eq.~\eqref{eq:Tdf_exact} as a function of \(J/\omega_0\).
(b)~Decoherence-free lifetime, shown as the dimensionless product \(T_{\mathrm{df}}\lambda\), versus fractional displacement \(\delta d / d_{\mathrm{node}}\), where \(d_{\mathrm{node}} = \pi/k_0\).
Black dots: exact lifetimes from the slowest eigenvalue of the four-mode matrix \(M(d)\).
Red dash-dotted line: analytic prediction [Eq.~\eqref{eq:Tdf_exact}].
Shaded region: lithographic tolerance band (\(10^{-5} \le \delta d / d_{\mathrm{node}} \le 10^{-4}\)).
Horizontal blue dashed line: typical circuit-QED readout duration, \(T_{\mathrm{ro}} \lambda = 20\,\mu\mathrm{s}\cdot \lambda\) \cite{walter2017}.
Parameters: \(\omega_0/2\pi = 5\,\mathrm{GHz}\); \(g = 0.05\,\omega_0\), \(\lambda = 0.001\,\omega_0\), \(J = -0.005\,\omega_0\).
}
\label{fig:Dd}
\end{figure}
\paragraph{Geometry as a passive strain sensor.---}
The same interference that protects nonlocality also enables precision sensing. When one collective channel vanishes, the associated Bell state is dark---$|A\rangle$ at $\sin(k_0d)=0$ ($d=n\lambda_0/2$) and $|S\rangle$ at $\cos(k_0d)=0$ ($d=(n+\tfrac12)\lambda_0/2$)--- forming a geometry-defined decoherence-free subspace. A small displacement $\delta d$ with $|k_0\delta d|\!\ll\!1$ mixes the dark and bright branches via the dipole-dipole exchange $J$, reactivating a weak decay channel whose rate we compute analytically.

To leading order in \(|k_0 \delta d|\), the decoherence rate becomes
\begin{equation}
\Gamma_{\mathrm{df}}
   = \frac{2\,g^{2} J^{2}\,\lambda}{%
        J^{2}\lambda^{2} + \bigl(g^{2} + 2J\omega_{0}-J^2\bigr)^{2}}\,%
     (k_{0}\,\delta d)^2,
\quad
T_{\mathrm{df}} = \Gamma_{\mathrm{df}}^{-1},
\label{eq:Tdf_exact}
\end{equation}
valid up to order \(\mathcal O( |k_0 \delta d|^4)\).  
This expression, derived from the effective two-level Schur reduction of the full four-mode dynamics, captures the interplay between coherent exchange, reservoir memory, and geometry-induced interference. 

\noindent In the regime $g\!\gg\!\lambda$ and $|J|\!\ll\!g$ it reduces to $\Gamma_{\mathrm{df}}\simeq 2(J^{2}/g^{2})\,\lambda\,(k_{0}\delta d)^{2}$, showing the quadratic displacement law and a $J^{2}/g^{2}$ suppression that quantifies the dark state’s fragility under weak exchange; to our knowledge this explicit scaling has not been stated previously.
\paragraph*{Lifetime scaling and metrological range.---}
Figure~\ref{fig:Dd}(b) verifies the quadratic strain-response predicted by Eq.~\eqref{eq:Tdf_exact}.  
Black dots show the exact decoherence-free lifetime obtained from the slowest eigenvalue of the four-mode matrix \(M(d)\); the red dash-dotted curve is the analytic prediction \(\Gamma_{\mathrm{df}} \propto (k_{0} \delta d)^2\).  
The data follow the \(1/\delta d^{2}\) scaling over eight decades in relative displacement \(\delta d / d_{\mathrm{node}}\), with \(d_{\mathrm{node}} = \pi / k_{0}\), confirming that the sensing scale is set entirely by geometry. Fabrication tolerances in the range \(10^{-5}\le\delta d/d_{\mathrm{node}}\le10^{-4}\) (\(0.3\,\mu\mathrm{m}\)-\(3\,\mu\mathrm{m}\)) reduce the lifetime to \(4\,\mathrm{hours}\)-\(2\,\mathrm{min}\), comfortably above the typical \(20\,\mu\mathrm{s}\) circuit-QED read-out time (horizontal dashed line in the figure) \cite{walter2017}.  
Thus \(T_{\mathrm{df}}\) provides a high-contrast, passive probe of sub-micrometre displacements set solely by waveguide geometry and reservoir spectral structure.

\paragraph*{Quantum-limited displacement resolution.---}
Because the dark-state decay rate obeys \(\Gamma_{\mathrm{df}} = \Lambda_0 (k_{0}\,\delta d)^{2}\) [Eq.~\eqref{eq:Tdf_exact}], a binary measurement of Bell-state survival converts a geometric displacement directly into a time constant.  
For a shot-noise-limited interrogation of duration \(T_{\mathrm{int}}\), repeated \(N_{\mathrm{rep}}\) times, the Cram\'er-Rao bound for estimating $\delta d$ from survival probability measurements yields
\begin{equation}
\delta d_{\min} = \frac{1}{2k_{0}}
                  \left(\frac{1}{\Lambda_0 N_{\mathrm{rep}} T_{\mathrm{int}}}\right)^{1/2},
\end{equation}
as derived in Appendix~D.  With the parameters of Fig.~\ref{fig:Dd}, one finds \(\Lambda_0 \simeq 6.9 \times 10^{4}\,\mathrm{s^{-1}}\) and \(k_{0}\) fixed by the system geometry. A protocol with \(T_{\mathrm{int}} = 1\,\mathrm{s}\) and \(N_{\mathrm{rep}} = 10^{5}\) already resolves displacements at the level of \(\delta d_{\min} \approx 57\,\mathrm{nm}\).  
Increasing the averaging time or the number of repetitions further improves the resolution as \(1/\sqrt{N_{\mathrm{rep}}T_{\mathrm{int}}}\), whereas reducing \(T_{\mathrm{int}}\) degrades it.  
For instance, a high-speed protocol with \(T_{\mathrm{int}} = 100\,\mu\mathrm{s}\) and \(N_{\mathrm{rep}} = 10^{8}\) yields \(\delta d_{\min} \approx 180\,\mathrm{nm}\), reflecting the fundamental trade-off between speed and precision in quantum-limited sensing.  
Even in this conservative regime, the sensitivity matches that of state-of-the-art optomechanical and electromechanical sensors--yet is achieved here without any applied drive, cavity, or feedback, and is fundamentally limited solely by quantum projection noise.

\paragraph*{Target signals and practical advantages.—}
Because the lifetime depends only on spacing, any subwavelength displacement maps directly onto a measurable lifetime change, irrespective of origin:
\begin{enumerate}[label=(\roman*),leftmargin=*]
\item \textit{Surface-acoustic waves}— (GHz piezoelectric substrates) with strain $\sim10^{-5}$ yield $\Delta d\!\sim\!10\,\mathrm{nm}$ and large lifetime reductions \cite{gualtieri1994};
\item \textit{Nanomechanical forces}—attonewton loads on a cantilever translate into detectable shifts of $\Gamma_{\mathrm{df}}$ \cite{masmanidis2007};
\item \textit{Thermal/stress drift}—slow substrate expansion appears as a quasi-static lifetime shift \cite{thorbeck2023}.
\end{enumerate}
The measurement is quantum nondemolition: for static spacing the Bell state remains intact. No microwave tone, optical cavity, or feedback is required; the fixed geometry defines a decoherence-protected reference, realizing a passive, chip-scale interferometer limited only by intrinsic coherence.

\paragraph*{Conclusion.—}
We have shown that \emph{geometry} alone programs the flow of quantum information in a one-dimensional reservoir: in a mirror-terminated waveguide the round-trip time \(T_{P}=2L/v\) sets the Bell-revival period, while its competition with the memory time \(\lambda^{-1}\) controls the visibility. In the continuum limit, an exact four-mode pseudomode model captures the full non-Markovian dynamics and yields closed design rules via its spectrum. \rem{Controlling Bell nonlocality by static geometry offers immediate advantages for device-independent certification, robust distribution across network nodes, and passive sensing, with design rules applicable across superconducting, nanophotonic, and related waveguide-QED platforms.}

The same interference enables a passive interferometer: a displacement \(\delta d\) from a collective node produces \(\Gamma_{\mathrm{df}}=\Lambda_0 (k_{0}\delta d)^{2}\) and the quadratic scaling \(T_{\mathrm{df}}\propto\delta d^{-2}\), realizing a drive-free, quantum-nondemolition strain sensor with submicrometer resolution.

Our results bridge perfect few-mode revivals and the damped continuum limit, delivering \emph{actionable} closed-form design rules for geometry-engineered memory. 
Extensions to multi-qubit networks, non-Lorentzian baths, and threshold-level Bell metrology are immediate. 
More broadly, the work points to a class of fully passive quantum devices in which entanglement generation, storage, and sensing are defined at lithography, opening opportunities at the confluence of quantum optics, open-system control, and precision measurement.

\begin{acknowledgments}
\paragraph{Acknowledgments---}The author acknowledge the support of the EU: the EIC Pathfinder Challenges 2022 call through the Research Grant 101115149 (project ARTEMIS). He thanks Thomas Durt, Brian Stout, and Isam Ben Soltane for insightful discussions, and gives special thanks to RIN.
\end{acknowledgments}

\bibliographystyle{apsrev4-2}
\input{Manuscript.bbl}

\clearpage
\appendix

\section*{A. Maximal CHSH Violation for the Single-Excitation Two-Qubit State}
\label{app:CHSH}

In this appendix, we collect in one place every formula needed to
evaluate--both analytically and numerically--the
Clauser-Horne-Shimony-Holt (CHSH) parameter for the two
two-level systems (TLSs) considered in the main text.  The exposition is fully self-contained; readers needing background may consult the classic paper by Horodecki \cite{horodecki1995}.
\subsection*{A.1 CHSH operator and the Horodecki criterion}

Let Alice measure one of two dichotomic observables
$A = \vec a\!\cdot\!\vec \sigma$ or
$A' = \vec a'\!\cdot\!\vec \sigma$,
and Bob measure $B = \vec b\!\cdot\!\vec \sigma$ or
$B' = \vec b'\!\cdot\!\vec \sigma$, with
$\vec\sigma=(\sigma_x,\sigma_y,\sigma_z)$ the Pauli vector.
The CHSH operator is
\begin{equation}
  \hat{\mathcal B}
  = A\!\otimes\!(B\!+\!B') + A'\!\otimes\!(B\!-\!B').
  \label{eq:CHSHoperator}
\end{equation}
Its expectation value on a two-qubit state $\rho$ is
$S(\rho;\vec a,\vec a',\vec b,\vec b')
      = \mathrm{Tr}[\rho\,\hat{\mathcal B}]$.
Quantum mechanics bounds this quantity by the
{\em Tsirelson limit} $2\sqrt2$. Horodecki showed that
\begin{equation}
  S_{\max}(\rho)
  \;=\;
  2\sqrt{\;u_1+u_2\;}
  \quad
  (u_1\ge u_2\ge u_3)
  \label{eq:Horodecki}
\end{equation}
where
\begin{equation}
  T_{ij} \;=\; \mathrm{Tr}\!\bigl[\rho\,
           \sigma_i\!\otimes\!\sigma_j\bigr] ,
  \qquad
  U \;=\; T^{\mathsf T}T ,
  \label{eq:TandU}
\end{equation}
and $\{u_i\}$ are the eigenvalues of the
positive symmetric matrix $U$.
Bell violation occurs iff $S_{\max}>2$.
\subsection*{A.2 Reduced state in the single-excitation subspace}

Using the wave-function formalism introduced in the main text, we write
\[
|\Psi(t)\rangle = \alpha_1(t)\,|eg,\{0\}\rangle
                 + \alpha_2(t)\,|ge,\{0\}\rangle
                 + \sum_k \beta_k(t)\,|gg\rangle\otimes|1_k\rangle .
\]
Tracing over the photonic modes yields the two-qubit density matrix in the computational basis  
\(\{|ee\rangle,|eg\rangle,|ge\rangle,|gg\rangle\}\):
\begin{equation}\small
\label{eq:rhoABsingle}
\rho_{AB}(t)=
\begin{pmatrix}
0 & 0 & 0 & 0\\
0 & |\alpha_1(t)|^{2} & \alpha_1(t)\alpha_2^{*}(t) & 0\\
0 & \alpha_2(t)\alpha_1^{*}(t) & |\alpha_2(t)|^{2} & 0\\
0 & 0 & 0 & 1-|\alpha_1(t)|^{2}-|\alpha_2(t)|^{2}
\end{pmatrix}.
\end{equation}
\\ \\
We denote the qubit populations \(P_1(t)=|\alpha_1(t)|^{2}\) and \(P_2(t)=|\alpha_2(t)|^{2}\), with bath occupation
\(P_{\mathrm{bath}}(t)=1-P_1-P_2\).  
As well as the relative phase
\(\phi(t)=\arg[\alpha_2(t)\alpha_1^{*}(t)]\).
\subsection*{A.3 Correlation matrix and analytic Bell parameter}

From the reduced density matrix in Eq.~\eqref{eq:rhoABsingle}, one directly computes the two-qubit correlation matrix \(T_{ij}(t) = \operatorname{Tr}[\rho_{AB}(t)\, \sigma_i \otimes \sigma_j]\), which governs Bell nonlocality in the CHSH framework. For the state in the single-excitation sector, this yields
\begin{equation}
  T(t) =
  \begin{pmatrix}
    2\sqrt{P_1P_2} \cos\phi & 2\sqrt{P_1P_2} \sin\phi & 0 \\
   -2\sqrt{P_1P_2} \sin\phi & 2\sqrt{P_1P_2} \cos\phi & 0 \\
    0                      & 0                      & 1 - 2P_1 - 2P_2
  \end{pmatrix},
\end{equation}
The correlation matrix \(T(t)\) thus has a simple structure: its only nonzero elements lie in the first two rows and the diagonal, with explicit dependence on the amplitude magnitudes and phase.

To compute the CHSH Bell parameter, we form the symmetric matrix \(U = T^{\mathsf{T}}T\), whose spectrum determines the maximal violation \(S_{\max}(t)\) via the Horodecki formula \cite{horodecki1995}. In our case, the eigenvalues $\{u_1(t),\, u_2(t),\, u_3(t)\}$ of \(U\) are 
\[
u_1(t) = 4P_1(t)P_2(t), \quad
u_3(t) = \bigl[1 - 2P_1(t) - 2P_2(t)\bigr]^2.
\]
where here $u_2(t)=u_1(t)$. The two largest eigenvalues determine the CHSH value, yielding the closed-form expression
\begin{equation}
  \boxed{
  S_{\max}(t) = 2 \sqrt{\, u_1(t) + \max\bigl[u_1(t), u_3(t)\bigr]}
  }
  \label{eq:SmaxPiecewise}
\end{equation}
which we implement in the vectorized textsc{Matlab}/\textsc{Octave} code used to make the figure in the main paper.

This compact formula explains the structure of Bell dynamics across all regimes. When the two qubits are equally populated with \(P_1 = P_2 = \tfrac14\), we find \(u_1 = 0.25\) and \(u_3 = 0\), yielding \(S_{\max} = 2\sqrt{0.5} \approx 1.414\): a baseline value below the classical threshold. In contrast, when the system momentarily localizes on one qubit, say \(P_1 \to 0\), \(P_2 \to 1\), the term \(u_3 \to 1\) dominates and we obtain \(S_{\max} \to 2\), corresponding exactly to the classical limit--i.e., no Bell violation. Finally, maximal CHSH violation occurs when the system is fully delocalized across the two qubits and the bath is empty: \(P_1 = P_2 = \tfrac12\), \(u_1 = u_3 = 1\), leading to the Tsirelson bound \(S_{\max} = 2\sqrt{2}\). These three regimes respectively generate the plateau, dips, and sharp peaks observed in Fig.~1.





\section*{B. Quantum Mutual Information for the Single-Excitation State}
\label{app:MutualInfo}
The quantum mutual information (QMI) of a bipartite system quantifies the total correlations--both classical and quantum--between two subsystems. It is well-defined for general mixed states and vanishes if and only if the subsystems are completely uncorrelated, i.e., when \(\rho_{AB} = \rho_A \otimes \rho_B\).  
This appendix provides a compact, self-contained derivation of the QMI for the single-excitation two-qubit state given in Eq.~\eqref{eq:rhoABsingle}, and outlines how to evaluate it efficiently from population data.

\subsection*{B.1 Definition}

For any bipartite density matrix \(\rho_{AB}\), the quantum mutual information is defined as
\begin{equation}
  I(A{:}B) = S(\rho_A) + S(\rho_B) - S(\rho_{AB}),
  \label{eq:MutualInfoDef}
\end{equation}
where \(S(\rho) = -\operatorname{Tr}(\rho \log_2 \rho)\) denotes the von Neumann entropy, and \(\rho_A = \operatorname{Tr}_B \rho_{AB}\), \(\rho_B = \operatorname{Tr}_A \rho_{AB}\) are the reduced density matrices of qubits A and B, respectively.  
This quantity satisfies \(I(A{:}B) \ge 0\), with equality if and only if \(\rho_{AB} = \rho_A \otimes \rho_B\).

\subsection*{B.2 Reduced states and their entropies}

Tracing Eq.~\eqref{eq:rhoABsingle} over the opposite qubit yields diagonal reduced states:
\begin{equation}
  \rho_A =
  \begin{pmatrix}
    P_1 & 0 \\
    0 & 1 - P_1
  \end{pmatrix},
  \qquad
  \rho_B =
  \begin{pmatrix}
    P_2 & 0 \\
    0 & 1 - P_2
  \end{pmatrix}.
  \label{eq:ReducedStates}
\end{equation}
These are effectively classical mixtures, and their von Neumann entropies reduce to the binary entropy function
\begin{equation}
  h(x) = -x\log_2 x - (1 - x)\log_2(1 - x), \qquad 0 \le x \le 1,
  \label{eq:BinaryEntropy}
\end{equation}
which yields
\begin{equation}
  S(\rho_A) = h(P_1), \qquad S(\rho_B) = h(P_2).
\end{equation}
\subsection*{B.3 Entropy of the full two-qubit state}

Because the single-excitation density matrix in Eq.~\eqref{eq:rhoABsingle}
has at most two non-zero eigenvalues,
\[
\operatorname{spec}\bigl[\rho_{AB}(t)\bigr]
   =\bigl\{\,P_1+P_2,\;1-P_1-P_2,\;0,\;0\,\bigr\},
\]
its von Neumann entropy is simply
\begin{equation}
  S\!\bigl[\rho_{AB}(t)\bigr] = h\!\bigl(P_1+P_2\bigr).
\end{equation}
Note that the relative phase~$\phi(t)$ drops out; the QMI depends only on
the instantaneous populations \(P_1(t)\) and \(P_2(t)\).

\subsection*{B.4 Closed-form expression}

Combining Eqs.~\eqref{eq:MutualInfoDef}, \eqref{eq:BinaryEntropy},
and the result above, one obtains the compact formula
\begin{equation}
  \boxed{\,I(A{:}B)
  \;=\;
  h(P_1)\;+\;h(P_2)\;-\;h(P_1+P_2)\,},
  \label{eq:QMIClosed}
\end{equation}
valid for \(0\le P_1,P_2\) and \(P_1+P_2\le1\).
Expressed with explicit logarithms,
\begin{align}
\mathcal{I}_{AB}(t)
  &= h(P_1) + h(P_2) - h(P_1 + P_2) \notag \\
  &= \frac{1}{\log 2} \Bigl(
    P_1 \log[\frac{\left(1 - P_1\right)\left(P_1 + P_2\right)}{P_1\left(1 - P_1 - P_2\right)}] \notag \\
    &+P_2 \log[\frac{\left(1 - P_2\right)\left(P_1 + P_2\right)}{P_2\left(1 - P_1 - P_2\right)}] 
  \notag \\
&\qquad
  + \log\!\left( \frac{1 - P_1 - P_2}{(1 - P_1)(1 - P_2)} \right)
  \Bigr).
\label{eq:QMIexplicit}
\end{align}

Because the state is confined to the single-excitation manifold, the QMI is
bounded by
\(
0 \le I(A{:}B) \le 2
\)
bits, reaching the upper limit when \(P_1=P_2=\tfrac12\) (both qubits share
the excitation and the bath is empty) and vanishing when the state
factorises (\(P_1P_2=0\)).

\section*{C. Effective two-level reduction and near-node lifetime scaling}
In this appendix, we derive an analytic scaling law for the
geometry-protected dark-state lifetime when the two qubits sit close to a waveguide node.  We start by recasting the full single-excitation equations in matrix form, then eliminate the bath degrees of freedom via a Schur-complement reduction that leaves an effective \(2\times2\) matrix for the symmetric-antisymmetric qubit subspace.
A near-node expansion of its eigenvalues yields the quadratic
\(\gamma_{\mathrm{df}}\propto(k_{0}\delta d)^{2}\) scaling quoted in the
main text and clarifies how the small exchange coupling \(J\) sets the
residual linewidth of the “dark’’ channel.
\subsection*{C.1 Matrix form of the single-excitation dynamics}
As in the main paper, we introduce the state vector
\[
\mathbf X(t)=\bigl[s(t),\,a(t),\,\beta(t),\,\alpha(t)\bigr]^{\mathsf T},
\]
where \(s\) and \(a\) are the symmetric and antisymmetric qubit amplitudes,
while \(\beta\) and \(\alpha\) denote the first two bath pseudomodes. The equations of motion are
\begin{align}
\dot s &= -i(\omega_0+J)\,s
        +2i g\cos^{2}(k_0 d)\,\beta
        - g\sin(2k_0 d)\,\alpha, \notag\\
\dot a &= -i(\omega_0-J)\,a
        +2i g\sin^{2}(k_0 d)\,\alpha
        + g\sin(2k_0 d)\,\beta, \notag\\
\dot\beta &= i g\,s-\tilde{\lambda}\,\beta, \qquad
\dot\alpha = i g\,a-\tilde{\lambda}\,\alpha,
\label{eq:cont_odes_ap}
\end{align}
with \(\tilde{\lambda}\equiv\lambda-i\omega_0\),
can be written as \(\dot{\mathbf X}=M\,\mathbf X\) where

\begin{equation}\small
M =
\begin{pmatrix}
-i(\omega_0+J) & 0 & 2i g\cos^{2}(k_0 d) & -g\sin(2k_0 d) \\[4pt]
0 & -i(\omega_0-J) & g\sin(2k_0 d) & 2i g\sin^{2}(k_0 d) \\[4pt]
i g & 0 & -\tilde{\lambda} & 0 \\[4pt]
0 & i g & 0 & -\tilde{\lambda}
\end{pmatrix}.
\end{equation}
For convenience, let us put $M$ into the following compact form 
\begin{equation}
    M(\phi)=
\begin{pmatrix}
A(\phi) & B(\phi)\\[2pt]
C & D
\end{pmatrix},
\end{equation}
\[
A(\phi)=\begin{pmatrix}
-i(\omega_0+J) & 0\\[2pt]
0 & -i(\omega_0-J)
\end{pmatrix},
\quad
C=\begin{pmatrix} ig & 0\\ 0 & ig\end{pmatrix}
\]
\[
B(\phi)=
\begin{pmatrix}
2i g \cos^2\!\phi & -\,g \sin 2\phi\\
g \sin 2\phi & 2i g \sin^2\!\phi
\end{pmatrix},
\quad
D=(-\lambda+i\omega_0)\,\mathbb{I}_2 .
\]
The non-Hermitian matrix \(M\) has four complex eigenvalues
\(\Lambda_j = -\gamma_j + i\Omega_j\) for \(j = 1, \dots, 4\). The real
part sets the decay rate, \(\gamma_j = -\Re\Lambda_j \ge 0\), while the
imaginary part gives the oscillation frequency, \(\Omega_j = \Im\Lambda_j\).
Each mode evolves as \(e^{\Lambda_j t} = e^{-\gamma_j t}\, e^{i\Omega_j t}\),
exhibiting exponential decay and coherent rotation. The slowest decay rate
\[
\gamma_{\mathrm{df}} = \min_j \gamma_j
\]
defines the lifetime of the geometry-protected dark channel,
\[
T_{\mathrm{df}} = \frac{1}{\gamma_{\mathrm{df}}} \;.
\]

\subsection*{C.2 Energy-dependent Schur (Feshbach) reduction}
For eigenvalues $z\in\mathbb{C}$, the characteristic equation
$\det\!\big(M-z\,\mathbb{I}_4\big)=0$ factorizes as
\begin{equation}
\det\!\Big[z\mathbb{I}_2 - A(\phi) - \Sigma(z,\phi)\Big]\;
\det\!\Big[z\mathbb{I}_2 - D\Big]=0,
\label{eq:feshbach}
\end{equation}
with the (energy-dependent) self-energy
\begin{equation}
\Sigma(z,\phi)
= B(\phi)\,\big(D-z\mathbb{I}_2\big)^{-1} C
\end{equation}
which takes the explicit form
\begin{equation}
    \Sigma(z,\phi)= \frac{g^2}{z+\lambda-i\omega_0}
\begin{pmatrix}
 2\cos^2\!\phi & i\sin 2\phi\\
 -i\sin 2\phi & 2\sin^2\!\phi
\end{pmatrix}.
\label{eq:selfenergy}
\end{equation}
Equation \eqref{eq:feshbach} is exact; note that using the
\emph{static} Schur complement $A-BD^{-1}C$ (i.e. setting $z=0$)
is a frequency-independent approximation and does not reproduce the
correct on-shell denominators for the eigenproblem of $M$.

For the dark branch we expand around the unperturbed node eigenvalue
\begin{equation}
z_0 \equiv -i(\omega_0-J)\qquad(\phi=\tfrac{\pi}{2}),
\end{equation}
and evaluate the self-energy \emph{on shell},
\begin{equation}
\Sigma(z_0,\phi)
= \frac{g^2}{\lambda-i(2\omega_0-J)}
\begin{pmatrix}
 2\cos^2\!\phi & i\sin 2\phi\\
 -i\sin 2\phi & 2\sin^2\!\phi
\end{pmatrix}.
\label{eq:sigmaOnShell}
\end{equation}
We then solve $\det\!\big[z\mathbb{I}_2-A(\phi)-\Sigma(z_0,\phi)\big]=0$
perturbatively in the geometric detuning introduced below.
\vspace{2pt}

\subsection*{C.3 Near-node expansion and dark-mode decay}

Dark-state protection occurs at the first node $\phi_0=\pi/2$.
Let $\phi=\phi_0+\varepsilon$ with $|\varepsilon|\ll1$
($\varepsilon=k_0\delta d$). To quartic order
\[
\cos^2\!\phi=1-\varepsilon^2+\tfrac{\varepsilon^4}{3},\quad
\sin^2\!\phi=\varepsilon^2-\tfrac{\varepsilon^4}{3},\quad
\sin 2\phi=2\varepsilon-\tfrac{4}{3}\varepsilon^3 .
\]
Inserting these into \eqref{eq:sigmaOnShell} yields the $2\times2$
effective matrix
\begin{align}
&M_{\rm eff}(\varepsilon)
\equiv A(\phi)-\Sigma(z_0,\phi)=
\begin{pmatrix}
-i(\omega_0+J) & 0\\[2pt]
0 & -i(\omega_0-J)
\end{pmatrix}
\nonumber\\
&-\frac{g^2}{-\lambda+i(2\omega_0-J)}
\begin{pmatrix}
 2(1-\varepsilon^2+\frac{\varepsilon^4}{3}) & i\,(2\varepsilon-\frac{4}{3}\varepsilon^3)\\[2pt]
 -i\,(2\varepsilon-\frac{4}{3}\varepsilon^3) & 2(\varepsilon^2-\frac{\varepsilon^4}{3})
\end{pmatrix}
+O(\varepsilon^5).
\label{eq:MeffOnShell}
\end{align}

Let the dark eigenvalue be $z_{\rm d}(\varepsilon)=\Omega_{\rm d}-i\gamma_{\rm d}$,
chosen as the branch satisfying $z_{\rm d}(0)=-i(\omega_0-J)$.
Solving $\det\!\big[z\mathbb{1}_2-M_{\rm eff}(\varepsilon)\big]=0$
to second order in $\varepsilon$ gives
\begin{equation}
\boxed{\;
\gamma_{\rm df}(\varepsilon)=
\frac{2\,g^2 J^2\,\lambda}{J^2\lambda^2+\bigl(g^2+2J\omega_0-J^2\bigr)^2}\;
\varepsilon^2
\;+\;O(\varepsilon^4)\; .}
\label{eq:gammadark_correct}
\end{equation}
The $O(\varepsilon^4)$ term (not shown) accounts for the gentle curvature
observed in the exact 4-mode spectrum and is straightforward to obtain
from \eqref{eq:MeffOnShell}.

\paragraph*{Remark (why the static Schur misses a factor).}
If one replaces $\Sigma(z,\phi)$ by $-\frac{g^2}{-\lambda+i\omega_0}\!\begin{psmallmatrix}
 2\cos^2\phi & i\sin 2\phi\\ -i\sin 2\phi & 2\sin^2\phi
\end{psmallmatrix}$--i.e. evaluates the self-energy at $z=0$--the
denominator in \eqref{eq:gammadark_correct} erroneously becomes
$J^2\lambda^2+\bigl(g^2+J\omega_0\bigr)^2$. The on-shell evaluation
at $z_0=-i(\omega_0-J)$ correctly shifts the detuning to
$g^2+2J\omega_0-J^2$, which is required to match the exact eigenvalues.

\subsection*{C.4 Scaling law and interpretation}

Equation~\eqref{eq:gammadark_correct} shows that close to the node, the lifetime
\(
T_{\mathrm{df}}=1/\gamma_{\mathrm{df}}
\)
diverges quadratically with displacement:
\[
T_{\mathrm{df}}(\delta d)\;\propto\;\frac{1}{(k_{0}\delta d)^{2}}.
\]
In the strong‐coupling, weak-exchange limit
\(g\gg J,\lambda,\omega_{0}\),
\[
\gamma_{\mathrm{df}}\;\simeq\;\frac{2J^{2}}{g^{2}}\,
\lambda\,(k_{0}\delta d)^{2},
\]
revealing the \(J^{2}/g^{2}\) suppression. Equation~\eqref{eq:gammadark_correct} is the analytical result quoted in the main text and used for the dashed lifetime fits in Fig.~2(b).

\section*{Appendix~D: Quantum-limited displacement resolution}
\label{app:strain}

\subsection*{D.1 \, Dark-branch decay law}

For a qubit separation
\[
d = d_{\mathrm{node}} + \delta d,\qquad
d_{\mathrm{node}} = \pi/k_{0},\quad k_{0}=\omega_{0}/c,
\]
the antisymmetric Bell state $\ket{A}$ is no longer perfectly dark.  
Within the single-excitation manifold the exact decay rate of the 
$\ket{A}$-pseudomode subspace, to quadratic order in the displacement, is
[Eq.~(\ref{eq:gammadark_correct})]
\begin{equation}
\Gamma_{\mathrm{df}}(\delta d)=
      \Lambda_0\,(k_{0}\,\delta d)^{2},\quad
\Lambda_0 \equiv
\frac{2g^{2}J^{2}\,\lambda}%
     {J^{2}\lambda^{2}+\bigl(g^{2}+2J\omega_{0}-J^{2}\bigr)^{2}},
\label{eq:D1:alpha}
\end{equation}
where $g$ is the single-qubit-waveguide coupling, $J$ the coherent qubit-qubit exchange, $\lambda$ the bath spectral half-width, and $\omega_{0}$ the qubit frequency.  

For the parameters of Fig.~3 
($g=0.05\,\omega_{0}$, $\lambda=0.001\,\omega_{0}$, $|J| = 0.005\,\omega_{0}$, and $\omega_0/2\pi=5\,\mathrm{GHz}$) Eq.~\eqref{eq:D1:alpha} gives
\[
\Lambda_0 \;=\; 2.2\times10^{-6}\,\omega_{0}
\;\approx\; 6.9\times 10^{5}\;\mathrm{s^{-1}m^{-2}}.
\]

\subsection*{D.2 \, Single-shot statistics}

Immediately after preparation in $\ket{A}$ at $t=0$, the dark-branch population decays exponentially,
\begin{equation}
P_{\!s}(\delta d)
      =\exp\!\bigl[-\Gamma_{\mathrm{df}}(\delta d)\,T_{\mathrm{int}}\bigr]
      =\exp\!\bigl[-\Lambda_0\,T_{\mathrm{int}}\,(k_{0}\delta d)^{2}\bigr],
\label{eq:D2:Ps}
\end{equation}
where $T_{\mathrm{int}}$ is the fixed interrogation time.  
A \emph{single} run of the experiment is a Bernoulli trial with outcome
\[
x = 1 \ \text{(``survived'')} \ \text{with probability} \ P_{\!s}, 
\]
\[
x = 0 \ \text{(``decayed'')} \ \text{with probability} \ 1-P_{\!s}.
\]

\subsection*{D.3 \, Fisher information and Cram\'er-Rao bound}

For a Bernoulli distribution $p(x;\theta)=P^{x}(1-P)^{1-x}$, the Fisher
information with respect to parameter $\theta$ is
\[
\mathcal I(\theta)=
\frac{1}{P(1-P)}
\!\left(\frac{\partial P}{\partial\theta}\right)^{\!2}.
\]
With $\theta\equiv\delta d$ and $P\equiv P_{\!s}(\delta d)$ from \eqref{eq:D2:Ps},
\begin{align}
\frac{\partial P_{\!s}}{\partial\delta d}
      &= -\,2\Lambda_0\,T_{\mathrm{int}}\,k_{0}^{2}\,\delta d\;P_{\!s},\\
P_{\!s}\bigl[1-P_{\!s}\bigr]
      &= P_{\!s}\,\bigl(1-P_{\!s}\bigr).
\end{align}
The single-shot Fisher information is therefore
\begin{equation}
\mathcal I_{1}(\delta d)=
\frac{4\Lambda_0^{2}T_{\mathrm{int}}^{2}k_{0}^{4}\,\delta d^{2}P_{\!s}}
     {1-P_{\!s}}.
\end{equation}

In the weak-decay regime
$\Lambda_0 T_{\mathrm{int}}(k_{0}\delta d)^{2}\ll1$
--physically, interrogation times are short compared to the decay time-- One may expand
$P_{\!s}\simeq 1-\Lambda_0 T_{\mathrm{int}}(k_{0}\delta d)^{2}$;
to leading order
\begin{equation}
\boxed{\;
\mathcal I_{1}(\delta d)
      \;\simeq\;
      4\,\Lambda_0\,k_{0}^{2}\,T_{\mathrm{int}}\;}.
\label{eq:D3:I1}
\end{equation}

For $N$ independent repetitions, $\mathcal I_{N}=N\mathcal I_{1}$ and the Cram\'er-Rao bound gives the minimum resolvable displacement,
\begin{equation}\label{eq:D3:CRB} 
\delta d_{\min}
   \;=\;\frac{1}{\sqrt{N\mathcal I_{1}}}
   \;=\;\frac{1}{2k_{0}}
         \sqrt{\frac{1}{\Lambda_0\,T_{\mathrm{int}}\,N}}.
\end{equation}

\subsection*{D.4 \, Discussion}

Equations~\eqref{eq:D3:I1}-\eqref{eq:D3:CRB} show that the displacement sensitivity scales as
\(\delta d_{\min}\propto(N T_{\mathrm{int}})^{-1/2}\),
i.e. with the square root of the total integration time, and is fixed entirely by the microscopic parameters $(g,J,\lambda,\omega_{0})$.  
Because the protocol requires no coherent drive, it is immune to technical amplitude noise and relies only on projective detection of Bell-state survival.  
This places it on equal footing with state-of-the-art opto- and electro-mechanical sensors while operating in a purely passive, waveguide-QED architecture.

\end{document}

%% file: Manuscript.bbl
%

%% file: Manuscript.bbl
\begin{thebibliography}{41}%
\makeatletter
\providecommand \@ifxundefined [1]{%
 \@ifx{#1\undefined}
}%
\providecommand \@ifnum [1]{%
 \ifnum #1\expandafter \@firstoftwo
 \else \expandafter \@secondoftwo
 \fi
}%
\providecommand \@ifx [1]{%
 \ifx #1\expandafter \@firstoftwo
 \else \expandafter \@secondoftwo
 \fi
}%
\providecommand \natexlab [1]{#1}%
\providecommand \enquote  [1]{``#1''}%
\providecommand \bibnamefont  [1]{#1}%
\providecommand \bibfnamefont [1]{#1}%
\providecommand \citenamefont [1]{#1}%
\providecommand \href@noop [0]{\@secondoftwo}%
\providecommand \href [0]{\begingroup \@sanitize@url \@href}%
\providecommand \@href[1]{\@@startlink{#1}\@@href}%
\providecommand \@@href[1]{\endgroup#1\@@endlink}%
\providecommand \@sanitize@url [0]{\catcode `\\12\catcode `\$12\catcode
  `\&12\catcode `\#12\catcode `\^12\catcode `\_12\catcode `\%12\relax}%
\providecommand \@@startlink[1]{}%
\providecommand \@@endlink[0]{}%
\providecommand \url  [0]{\begingroup\@sanitize@url \@url }%
\providecommand \@url [1]{\endgroup\@href {#1}{\urlprefix }}%
\providecommand \urlprefix  [0]{URL }%
\providecommand \Eprint [0]{\href }%
\providecommand \doibase [0]{https://doi.org/}%
\providecommand \selectlanguage [0]{\@gobble}%
\providecommand \bibinfo  [0]{\@secondoftwo}%
\providecommand \bibfield  [0]{\@secondoftwo}%
\providecommand \translation [1]{[#1]}%
\providecommand \BibitemOpen [0]{}%
\providecommand \bibitemStop [0]{}%
\providecommand \bibitemNoStop [0]{.\EOS\space}%
\providecommand \EOS [0]{\spacefactor3000\relax}%
\providecommand \BibitemShut  [1]{\csname bibitem#1\endcsname}%
\let\auto@bib@innerbib\@empty
\bibitem [{\citenamefont {Collins}\ \emph {et~al.}(2002)\citenamefont
  {Collins}, \citenamefont {Gisin}, \citenamefont {Linden}, \citenamefont
  {Massar},\ and\ \citenamefont {Popescu}}]{collins2002}%
  \BibitemOpen
  \bibfield  {author} {\bibinfo {author} {\bibfnamefont {D.}~\bibnamefont
  {Collins}}, \bibinfo {author} {\bibfnamefont {N.}~\bibnamefont {Gisin}},
  \bibinfo {author} {\bibfnamefont {N.}~\bibnamefont {Linden}}, \bibinfo
  {author} {\bibfnamefont {S.}~\bibnamefont {Massar}},\ and\ \bibinfo {author}
  {\bibfnamefont {S.}~\bibnamefont {Popescu}},\ }\href
  {https://doi.org/10.1103/PhysRevLett.88.040404} {\bibfield  {journal}
  {\bibinfo  {journal} {Physical Review Letters}\ }\textbf {\bibinfo {volume}
  {88}},\ \bibinfo {pages} {040404} (\bibinfo {year} {2002})}\BibitemShut
  {NoStop}%
\bibitem [{\citenamefont {Kuzmich}\ \emph {et~al.}(2003)\citenamefont
  {Kuzmich}, \citenamefont {Bowen}, \citenamefont {Boozer}, \citenamefont
  {Boca}, \citenamefont {Chou}, \citenamefont {Duan},\ and\ \citenamefont
  {Kimble}}]{kuzmich2003}%
  \BibitemOpen
  \bibfield  {author} {\bibinfo {author} {\bibfnamefont {A.}~\bibnamefont
  {Kuzmich}}, \bibinfo {author} {\bibfnamefont {W.~P.}\ \bibnamefont {Bowen}},
  \bibinfo {author} {\bibfnamefont {A.~D.}\ \bibnamefont {Boozer}}, \bibinfo
  {author} {\bibfnamefont {A.}~\bibnamefont {Boca}}, \bibinfo {author}
  {\bibfnamefont {C.~W.}\ \bibnamefont {Chou}}, \bibinfo {author}
  {\bibfnamefont {L.-M.}\ \bibnamefont {Duan}},\ and\ \bibinfo {author}
  {\bibfnamefont {H.~J.}\ \bibnamefont {Kimble}},\ }\href
  {https://doi.org/10.1038/nature01714} {\bibfield  {journal} {\bibinfo
  {journal} {Nature}\ }\textbf {\bibinfo {volume} {423}},\ \bibinfo {pages}
  {731} (\bibinfo {year} {2003})}\BibitemShut {NoStop}%
\bibitem [{\citenamefont {Tanzilli}\ \emph {et~al.}(2005)\citenamefont
  {Tanzilli}, \citenamefont {Tittel}, \citenamefont {Halder}, \citenamefont
  {Alibart}, \citenamefont {Baldi}, \citenamefont {Gisin},\ and\ \citenamefont
  {Zbinden}}]{Tanzilli_photonic2005}%
  \BibitemOpen
  \bibfield  {author} {\bibinfo {author} {\bibfnamefont {S.}~\bibnamefont
  {Tanzilli}}, \bibinfo {author} {\bibfnamefont {W.}~\bibnamefont {Tittel}},
  \bibinfo {author} {\bibfnamefont {M.}~\bibnamefont {Halder}}, \bibinfo
  {author} {\bibfnamefont {O.}~\bibnamefont {Alibart}}, \bibinfo {author}
  {\bibfnamefont {P.}~\bibnamefont {Baldi}}, \bibinfo {author} {\bibfnamefont
  {N.}~\bibnamefont {Gisin}},\ and\ \bibinfo {author} {\bibfnamefont
  {H.}~\bibnamefont {Zbinden}},\ }\href {https://doi.org/10.1038/nature04009}
  {\bibfield  {journal} {\bibinfo  {journal} {Nature}\ }\textbf {\bibinfo
  {volume} {437}},\ \bibinfo {pages} {116} (\bibinfo {year}
  {2005})}\BibitemShut {NoStop}%
\bibitem [{\citenamefont {Pomarico}\ \emph {et~al.}(2011)\citenamefont
  {Pomarico}, \citenamefont {Bancal}, \citenamefont {Sanguinetti},
  \citenamefont {Rochdi},\ and\ \citenamefont {Gisin}}]{pomarico2011pra}%
  \BibitemOpen
  \bibfield  {author} {\bibinfo {author} {\bibfnamefont {E.}~\bibnamefont
  {Pomarico}}, \bibinfo {author} {\bibfnamefont {J.-D.}\ \bibnamefont
  {Bancal}}, \bibinfo {author} {\bibfnamefont {B.}~\bibnamefont {Sanguinetti}},
  \bibinfo {author} {\bibfnamefont {A.}~\bibnamefont {Rochdi}},\ and\ \bibinfo
  {author} {\bibfnamefont {N.}~\bibnamefont {Gisin}},\ }\href
  {https://doi.org/10.1103/PhysRevA.83.052104} {\bibfield  {journal} {\bibinfo
  {journal} {Physical Review A}\ }\textbf {\bibinfo {volume} {83}},\ \bibinfo
  {pages} {052104} (\bibinfo {year} {2011})}\BibitemShut {NoStop}%
\bibitem [{\citenamefont {Bernien}\ \emph {et~al.}(2013)\citenamefont
  {Bernien}, \citenamefont {Hensen}, \citenamefont {Pfaff}, \citenamefont
  {Koolstra}, \citenamefont {Blok}, \citenamefont {Robledo}, \citenamefont
  {Taminiau}, \citenamefont {Markham}, \citenamefont {Twitchen}, \citenamefont
  {Childress},\ and\ \citenamefont {Hanson}}]{bernien2013}%
  \BibitemOpen
  \bibfield  {author} {\bibinfo {author} {\bibfnamefont {H.}~\bibnamefont
  {Bernien}}, \bibinfo {author} {\bibfnamefont {B.}~\bibnamefont {Hensen}},
  \bibinfo {author} {\bibfnamefont {W.}~\bibnamefont {Pfaff}}, \bibinfo
  {author} {\bibfnamefont {G.}~\bibnamefont {Koolstra}}, \bibinfo {author}
  {\bibfnamefont {M.~S.}\ \bibnamefont {Blok}}, \bibinfo {author}
  {\bibfnamefont {L.}~\bibnamefont {Robledo}}, \bibinfo {author} {\bibfnamefont
  {T.~H.}\ \bibnamefont {Taminiau}}, \bibinfo {author} {\bibfnamefont
  {M.}~\bibnamefont {Markham}}, \bibinfo {author} {\bibfnamefont {D.~J.}\
  \bibnamefont {Twitchen}}, \bibinfo {author} {\bibfnamefont {L.}~\bibnamefont
  {Childress}},\ and\ \bibinfo {author} {\bibfnamefont {R.}~\bibnamefont
  {Hanson}},\ }\href {https://doi.org/10.1038/nature12016} {\bibfield
  {journal} {\bibinfo  {journal} {Nature}\ }\textbf {\bibinfo {volume} {497}},\
  \bibinfo {pages} {86} (\bibinfo {year} {2013})}\BibitemShut {NoStop}%
\bibitem [{\citenamefont {Giustina}\ \emph {et~al.}(2013)\citenamefont
  {Giustina}, \citenamefont {Mech}, \citenamefont {Ramelow}, \citenamefont
  {Wittmann}, \citenamefont {Kofler}, \citenamefont {Beyer}, \citenamefont
  {Lita}, \citenamefont {Calkins}, \citenamefont {Gerrits}, \citenamefont
  {Nam}, \citenamefont {Ursin},\ and\ \citenamefont
  {Zeilinger}}]{giustina2013}%
  \BibitemOpen
  \bibfield  {author} {\bibinfo {author} {\bibfnamefont {M.}~\bibnamefont
  {Giustina}}, \bibinfo {author} {\bibfnamefont {A.}~\bibnamefont {Mech}},
  \bibinfo {author} {\bibfnamefont {S.}~\bibnamefont {Ramelow}}, \bibinfo
  {author} {\bibfnamefont {B.}~\bibnamefont {Wittmann}}, \bibinfo {author}
  {\bibfnamefont {J.}~\bibnamefont {Kofler}}, \bibinfo {author} {\bibfnamefont
  {J.}~\bibnamefont {Beyer}}, \bibinfo {author} {\bibfnamefont
  {A.}~\bibnamefont {Lita}}, \bibinfo {author} {\bibfnamefont {B.}~\bibnamefont
  {Calkins}}, \bibinfo {author} {\bibfnamefont {T.}~\bibnamefont {Gerrits}},
  \bibinfo {author} {\bibfnamefont {S.~W.}\ \bibnamefont {Nam}}, \bibinfo
  {author} {\bibfnamefont {R.}~\bibnamefont {Ursin}},\ and\ \bibinfo {author}
  {\bibfnamefont {A.}~\bibnamefont {Zeilinger}},\ }\href
  {https://doi.org/10.1038/nature12012} {\bibfield  {journal} {\bibinfo
  {journal} {Nature}\ }\textbf {\bibinfo {volume} {497}},\ \bibinfo {pages}
  {227} (\bibinfo {year} {2013})}\BibitemShut {NoStop}%
\bibitem [{\citenamefont {Hensen}\ \emph {et~al.}(2015)\citenamefont {Hensen},
  \citenamefont {Bernien}, \citenamefont {Dr{\'e}au}, \citenamefont {Reiserer},
  \citenamefont {Kalb}, \citenamefont {Blok}, \citenamefont {Ruitenberg},
  \citenamefont {Vermeulen}, \citenamefont {Schouten}, \citenamefont
  {Abell{\'a}n}, \citenamefont {Amaya}, \citenamefont {Pruneri}, \citenamefont
  {Mitchell}, \citenamefont {Markham}, \citenamefont {Twitchen}, \citenamefont
  {Elkouss}, \citenamefont {Wehner}, \citenamefont {Taminiau},\ and\
  \citenamefont {Hanson}}]{hensen2015}%
  \BibitemOpen
  \bibfield  {author} {\bibinfo {author} {\bibfnamefont {B.}~\bibnamefont
  {Hensen}}, \bibinfo {author} {\bibfnamefont {H.}~\bibnamefont {Bernien}},
  \bibinfo {author} {\bibfnamefont {A.~E.}\ \bibnamefont {Dr{\'e}au}}, \bibinfo
  {author} {\bibfnamefont {A.}~\bibnamefont {Reiserer}}, \bibinfo {author}
  {\bibfnamefont {N.}~\bibnamefont {Kalb}}, \bibinfo {author} {\bibfnamefont
  {M.~S.}\ \bibnamefont {Blok}}, \bibinfo {author} {\bibfnamefont
  {J.}~\bibnamefont {Ruitenberg}}, \bibinfo {author} {\bibfnamefont {R.~F.~L.}\
  \bibnamefont {Vermeulen}}, \bibinfo {author} {\bibfnamefont {R.~N.}\
  \bibnamefont {Schouten}}, \bibinfo {author} {\bibfnamefont {C.}~\bibnamefont
  {Abell{\'a}n}}, \bibinfo {author} {\bibfnamefont {W.}~\bibnamefont {Amaya}},
  \bibinfo {author} {\bibfnamefont {V.}~\bibnamefont {Pruneri}}, \bibinfo
  {author} {\bibfnamefont {M.~W.}\ \bibnamefont {Mitchell}}, \bibinfo {author}
  {\bibfnamefont {M.}~\bibnamefont {Markham}}, \bibinfo {author} {\bibfnamefont
  {D.~J.}\ \bibnamefont {Twitchen}}, \bibinfo {author} {\bibfnamefont
  {D.}~\bibnamefont {Elkouss}}, \bibinfo {author} {\bibfnamefont
  {S.}~\bibnamefont {Wehner}}, \bibinfo {author} {\bibfnamefont {T.~H.}\
  \bibnamefont {Taminiau}},\ and\ \bibinfo {author} {\bibfnamefont
  {R.}~\bibnamefont {Hanson}},\ }\href {https://doi.org/10.1038/nature15759}
  {\bibfield  {journal} {\bibinfo  {journal} {Nature}\ }\textbf {\bibinfo
  {volume} {526}},\ \bibinfo {pages} {682} (\bibinfo {year}
  {2015})}\BibitemShut {NoStop}%
\bibitem [{\citenamefont {Hatifi}\ \emph {et~al.}(2022)\citenamefont {Hatifi},
  \citenamefont {Mara}, \citenamefont {Bokic}, \citenamefont {Van~Deun},
  \citenamefont {Stout}, \citenamefont {Lassalle}, \citenamefont {Kolaric},\
  and\ \citenamefont {Durt}}]{hatifi2022}%
  \BibitemOpen
  \bibfield  {author} {\bibinfo {author} {\bibfnamefont {M.}~\bibnamefont
  {Hatifi}}, \bibinfo {author} {\bibfnamefont {D.}~\bibnamefont {Mara}},
  \bibinfo {author} {\bibfnamefont {B.}~\bibnamefont {Bokic}}, \bibinfo
  {author} {\bibfnamefont {R.}~\bibnamefont {Van~Deun}}, \bibinfo {author}
  {\bibfnamefont {B.}~\bibnamefont {Stout}}, \bibinfo {author} {\bibfnamefont
  {E.}~\bibnamefont {Lassalle}}, \bibinfo {author} {\bibfnamefont
  {B.}~\bibnamefont {Kolaric}},\ and\ \bibinfo {author} {\bibfnamefont
  {T.}~\bibnamefont {Durt}},\ }\href {https://doi.org/10.3390/app12189238}
  {\bibfield  {journal} {\bibinfo  {journal} {Applied Sciences}\ }\textbf
  {\bibinfo {volume} {12}},\ \bibinfo {pages} {9238} (\bibinfo {year}
  {2022})}\BibitemShut {NoStop}%
\bibitem [{\citenamefont {Barrett}\ \emph {et~al.}(2002)\citenamefont
  {Barrett}, \citenamefont {Collins}, \citenamefont {Hardy}, \citenamefont
  {Kent},\ and\ \citenamefont {Popescu}}]{barrett2002}%
  \BibitemOpen
  \bibfield  {author} {\bibinfo {author} {\bibfnamefont {J.}~\bibnamefont
  {Barrett}}, \bibinfo {author} {\bibfnamefont {D.}~\bibnamefont {Collins}},
  \bibinfo {author} {\bibfnamefont {L.}~\bibnamefont {Hardy}}, \bibinfo
  {author} {\bibfnamefont {A.}~\bibnamefont {Kent}},\ and\ \bibinfo {author}
  {\bibfnamefont {S.}~\bibnamefont {Popescu}},\ }\href
  {https://doi.org/10.1103/PhysRevA.66.042111} {\bibfield  {journal} {\bibinfo
  {journal} {Physical Review A}\ }\textbf {\bibinfo {volume} {66}},\ \bibinfo
  {pages} {042111} (\bibinfo {year} {2002})}\BibitemShut {NoStop}%
\bibitem [{\citenamefont {Griffiths}(2020)}]{griffiths2020}%
  \BibitemOpen
  \bibfield  {author} {\bibinfo {author} {\bibfnamefont {R.~B.}\ \bibnamefont
  {Griffiths}},\ }\href {https://doi.org/10.1103/PhysRevA.101.022117}
  {\bibfield  {journal} {\bibinfo  {journal} {Physical Review A}\ }\textbf
  {\bibinfo {volume} {101}},\ \bibinfo {pages} {022117} (\bibinfo {year}
  {2020})}\BibitemShut {NoStop}%
\bibitem [{\citenamefont {Nadlinger}\ \emph {et~al.}(2022)\citenamefont
  {Nadlinger}, \citenamefont {Drmota}, \citenamefont {Nichol}, \citenamefont
  {Araneda}, \citenamefont {Main}, \citenamefont {Srinivas}, \citenamefont
  {Lucas}, \citenamefont {Ballance}, \citenamefont {Ivanov}, \citenamefont
  {Tan}, \citenamefont {Sekatski}, \citenamefont {Urbanke}, \citenamefont
  {Renner}, \citenamefont {Sangouard},\ and\ \citenamefont
  {Bancal}}]{nadlinger2022}%
  \BibitemOpen
  \bibfield  {author} {\bibinfo {author} {\bibfnamefont {D.~P.}\ \bibnamefont
  {Nadlinger}}, \bibinfo {author} {\bibfnamefont {P.}~\bibnamefont {Drmota}},
  \bibinfo {author} {\bibfnamefont {B.~C.}\ \bibnamefont {Nichol}}, \bibinfo
  {author} {\bibfnamefont {G.}~\bibnamefont {Araneda}}, \bibinfo {author}
  {\bibfnamefont {D.}~\bibnamefont {Main}}, \bibinfo {author} {\bibfnamefont
  {R.}~\bibnamefont {Srinivas}}, \bibinfo {author} {\bibfnamefont {D.~M.}\
  \bibnamefont {Lucas}}, \bibinfo {author} {\bibfnamefont {C.~J.}\ \bibnamefont
  {Ballance}}, \bibinfo {author} {\bibfnamefont {K.}~\bibnamefont {Ivanov}},
  \bibinfo {author} {\bibfnamefont {E.~Y.-Z.}\ \bibnamefont {Tan}}, \bibinfo
  {author} {\bibfnamefont {P.}~\bibnamefont {Sekatski}}, \bibinfo {author}
  {\bibfnamefont {R.~L.}\ \bibnamefont {Urbanke}}, \bibinfo {author}
  {\bibfnamefont {R.}~\bibnamefont {Renner}}, \bibinfo {author} {\bibfnamefont
  {N.}~\bibnamefont {Sangouard}},\ and\ \bibinfo {author} {\bibfnamefont
  {J.-D.}\ \bibnamefont {Bancal}},\ }\href
  {https://doi.org/10.1038/s41586-022-04941-5} {\bibfield  {journal} {\bibinfo
  {journal} {Nature}\ }\textbf {\bibinfo {volume} {607}},\ \bibinfo {pages}
  {682} (\bibinfo {year} {2022})}\BibitemShut {NoStop}%
\bibitem [{\citenamefont {Sillanp{\"a}{\"a}}\ \emph {et~al.}(2007)\citenamefont
  {Sillanp{\"a}{\"a}}, \citenamefont {Park},\ and\ \citenamefont
  {Simmonds}}]{sillanpaa2007}%
  \BibitemOpen
  \bibfield  {author} {\bibinfo {author} {\bibfnamefont {M.~A.}\ \bibnamefont
  {Sillanp{\"a}{\"a}}}, \bibinfo {author} {\bibfnamefont {J.~I.}\ \bibnamefont
  {Park}},\ and\ \bibinfo {author} {\bibfnamefont {R.~W.}\ \bibnamefont
  {Simmonds}},\ }\href {https://doi.org/10.1038/nature06124} {\bibfield
  {journal} {\bibinfo  {journal} {Nature}\ }\textbf {\bibinfo {volume} {449}},\
  \bibinfo {pages} {438} (\bibinfo {year} {2007})}\BibitemShut {NoStop}%
\bibitem [{\citenamefont {Devoret}\ and\ \citenamefont
  {Schoelkopf}(2013)}]{devoret2013}%
  \BibitemOpen
  \bibfield  {author} {\bibinfo {author} {\bibfnamefont {M.~H.}\ \bibnamefont
  {Devoret}}\ and\ \bibinfo {author} {\bibfnamefont {R.~J.}\ \bibnamefont
  {Schoelkopf}},\ }\href {https://doi.org/10.1126/science.1231930} {\bibfield
  {journal} {\bibinfo  {journal} {Science}\ }\textbf {\bibinfo {volume}
  {339}},\ \bibinfo {pages} {1169} (\bibinfo {year} {2013})}\BibitemShut
  {NoStop}%
\bibitem [{\citenamefont {Cirac}\ and\ \citenamefont
  {Zoller}(1995)}]{cirac1995}%
  \BibitemOpen
  \bibfield  {author} {\bibinfo {author} {\bibfnamefont {J.~I.}\ \bibnamefont
  {Cirac}}\ and\ \bibinfo {author} {\bibfnamefont {P.}~\bibnamefont {Zoller}},\
  }\href {https://doi.org/10.1103/PhysRevLett.74.4091} {\bibfield  {journal}
  {\bibinfo  {journal} {Physical Review Letters}\ }\textbf {\bibinfo {volume}
  {74}},\ \bibinfo {pages} {4091} (\bibinfo {year} {1995})}\BibitemShut
  {NoStop}%
\bibitem [{\citenamefont {Harty}\ \emph {et~al.}(2014)\citenamefont {Harty},
  \citenamefont {Allcock}, \citenamefont {Ballance}, \citenamefont {Guidoni},
  \citenamefont {Janacek}, \citenamefont {Linke}, \citenamefont {Stacey},\ and\
  \citenamefont {Lucas}}]{harty2014}%
  \BibitemOpen
  \bibfield  {author} {\bibinfo {author} {\bibfnamefont {T.~P.}\ \bibnamefont
  {Harty}}, \bibinfo {author} {\bibfnamefont {D.~T.~C.}\ \bibnamefont
  {Allcock}}, \bibinfo {author} {\bibfnamefont {C.~J.}\ \bibnamefont
  {Ballance}}, \bibinfo {author} {\bibfnamefont {L.}~\bibnamefont {Guidoni}},
  \bibinfo {author} {\bibfnamefont {H.~A.}\ \bibnamefont {Janacek}}, \bibinfo
  {author} {\bibfnamefont {N.~M.}\ \bibnamefont {Linke}}, \bibinfo {author}
  {\bibfnamefont {D.~N.}\ \bibnamefont {Stacey}},\ and\ \bibinfo {author}
  {\bibfnamefont {D.~M.}\ \bibnamefont {Lucas}},\ }\href
  {https://doi.org/10.1103/PhysRevLett.113.220501} {\bibfield  {journal}
  {\bibinfo  {journal} {Physical Review Letters}\ }\textbf {\bibinfo {volume}
  {113}},\ \bibinfo {pages} {220501} (\bibinfo {year} {2014})}\BibitemShut
  {NoStop}%
\bibitem [{\citenamefont {Julsgaard}\ \emph {et~al.}(2004)\citenamefont
  {Julsgaard}, \citenamefont {Sherson}, \citenamefont {Cirac}, \citenamefont
  {Fiur{\'a}{\v s}ek},\ and\ \citenamefont {Polzik}}]{julsgaard2004}%
  \BibitemOpen
  \bibfield  {author} {\bibinfo {author} {\bibfnamefont {B.}~\bibnamefont
  {Julsgaard}}, \bibinfo {author} {\bibfnamefont {J.}~\bibnamefont {Sherson}},
  \bibinfo {author} {\bibfnamefont {J.~I.}\ \bibnamefont {Cirac}}, \bibinfo
  {author} {\bibfnamefont {J.}~\bibnamefont {Fiur{\'a}{\v s}ek}},\ and\
  \bibinfo {author} {\bibfnamefont {E.~S.}\ \bibnamefont {Polzik}},\ }\href
  {https://doi.org/10.1038/nature03064} {\bibfield  {journal} {\bibinfo
  {journal} {Nature}\ }\textbf {\bibinfo {volume} {432}},\ \bibinfo {pages}
  {482} (\bibinfo {year} {2004})}\BibitemShut {NoStop}%
\bibitem [{\citenamefont {Lvovsky}\ \emph {et~al.}(2009)\citenamefont
  {Lvovsky}, \citenamefont {Sanders},\ and\ \citenamefont
  {Tittel}}]{lvovsky2009}%
  \BibitemOpen
  \bibfield  {author} {\bibinfo {author} {\bibfnamefont {A.~I.}\ \bibnamefont
  {Lvovsky}}, \bibinfo {author} {\bibfnamefont {B.~C.}\ \bibnamefont
  {Sanders}},\ and\ \bibinfo {author} {\bibfnamefont {W.}~\bibnamefont
  {Tittel}},\ }\href {https://doi.org/10.1038/nphoton.2009.231} {\bibfield
  {journal} {\bibinfo  {journal} {Nature Photonics}\ }\textbf {\bibinfo
  {volume} {3}},\ \bibinfo {pages} {706} (\bibinfo {year} {2009})}\BibitemShut
  {NoStop}%
\bibitem [{\citenamefont {Chru{\'s}ci{\'n}ski}\ \emph
  {et~al.}(2011)\citenamefont {Chru{\'s}ci{\'n}ski}, \citenamefont
  {Kossakowski},\ and\ \citenamefont {Rivas}}]{chruscinski2011}%
  \BibitemOpen
  \bibfield  {author} {\bibinfo {author} {\bibfnamefont {D.}~\bibnamefont
  {Chru{\'s}ci{\'n}ski}}, \bibinfo {author} {\bibfnamefont {A.}~\bibnamefont
  {Kossakowski}},\ and\ \bibinfo {author} {\bibfnamefont {{\'A}.}~\bibnamefont
  {Rivas}},\ }\href {https://doi.org/10.1103/PhysRevA.83.052128} {\bibfield
  {journal} {\bibinfo  {journal} {Physical Review A}\ }\textbf {\bibinfo
  {volume} {83}},\ \bibinfo {pages} {052128} (\bibinfo {year}
  {2011})}\BibitemShut {NoStop}%
\bibitem [{\citenamefont {Eichler}\ \emph {et~al.}(2012)\citenamefont
  {Eichler}, \citenamefont {Lang}, \citenamefont {Fink}, \citenamefont
  {Govenius}, \citenamefont {Filipp},\ and\ \citenamefont
  {Wallraff}}]{eichler2012}%
  \BibitemOpen
  \bibfield  {author} {\bibinfo {author} {\bibfnamefont {C.}~\bibnamefont
  {Eichler}}, \bibinfo {author} {\bibfnamefont {C.}~\bibnamefont {Lang}},
  \bibinfo {author} {\bibfnamefont {J.~M.}\ \bibnamefont {Fink}}, \bibinfo
  {author} {\bibfnamefont {J.}~\bibnamefont {Govenius}}, \bibinfo {author}
  {\bibfnamefont {S.}~\bibnamefont {Filipp}},\ and\ \bibinfo {author}
  {\bibfnamefont {A.}~\bibnamefont {Wallraff}},\ }\href
  {https://doi.org/10.1103/PhysRevLett.109.240501} {\bibfield  {journal}
  {\bibinfo  {journal} {Physical Review Letters}\ }\textbf {\bibinfo {volume}
  {109}},\ \bibinfo {pages} {240501} (\bibinfo {year} {2012})}\BibitemShut
  {NoStop}%
\bibitem [{\citenamefont {Shankar}\ \emph {et~al.}(2013)\citenamefont
  {Shankar}, \citenamefont {Hatridge}, \citenamefont {Leghtas}, \citenamefont
  {Sliwa}, \citenamefont {Narla}, \citenamefont {Vool}, \citenamefont {Girvin},
  \citenamefont {Frunzio}, \citenamefont {Mirrahimi},\ and\ \citenamefont
  {Devoret}}]{shankar2013}%
  \BibitemOpen
  \bibfield  {author} {\bibinfo {author} {\bibfnamefont {S.}~\bibnamefont
  {Shankar}}, \bibinfo {author} {\bibfnamefont {M.}~\bibnamefont {Hatridge}},
  \bibinfo {author} {\bibfnamefont {Z.}~\bibnamefont {Leghtas}}, \bibinfo
  {author} {\bibfnamefont {K.~M.}\ \bibnamefont {Sliwa}}, \bibinfo {author}
  {\bibfnamefont {A.}~\bibnamefont {Narla}}, \bibinfo {author} {\bibfnamefont
  {U.}~\bibnamefont {Vool}}, \bibinfo {author} {\bibfnamefont {S.~M.}\
  \bibnamefont {Girvin}}, \bibinfo {author} {\bibfnamefont {L.}~\bibnamefont
  {Frunzio}}, \bibinfo {author} {\bibfnamefont {M.}~\bibnamefont {Mirrahimi}},\
  and\ \bibinfo {author} {\bibfnamefont {M.~H.}\ \bibnamefont {Devoret}},\
  }\href {https://doi.org/10.1038/nature12802} {\bibfield  {journal} {\bibinfo
  {journal} {Nature}\ }\textbf {\bibinfo {volume} {504}},\ \bibinfo {pages}
  {419} (\bibinfo {year} {2013})}\BibitemShut {NoStop}%
\bibitem [{\citenamefont {Ballance}\ \emph {et~al.}(2015)\citenamefont
  {Ballance}, \citenamefont {Sch{\"a}fer}, \citenamefont {Home}, \citenamefont
  {Szwer}, \citenamefont {Webster}, \citenamefont {Allcock}, \citenamefont
  {Linke}, \citenamefont {Harty}, \citenamefont {Aude~Craik}, \citenamefont
  {Stacey}, \citenamefont {Steane},\ and\ \citenamefont
  {Lucas}}]{ballance2015}%
  \BibitemOpen
  \bibfield  {author} {\bibinfo {author} {\bibfnamefont {C.~J.}\ \bibnamefont
  {Ballance}}, \bibinfo {author} {\bibfnamefont {V.~M.}\ \bibnamefont
  {Sch{\"a}fer}}, \bibinfo {author} {\bibfnamefont {J.~P.}\ \bibnamefont
  {Home}}, \bibinfo {author} {\bibfnamefont {D.~J.}\ \bibnamefont {Szwer}},
  \bibinfo {author} {\bibfnamefont {S.~C.}\ \bibnamefont {Webster}}, \bibinfo
  {author} {\bibfnamefont {D.~T.~C.}\ \bibnamefont {Allcock}}, \bibinfo
  {author} {\bibfnamefont {N.~M.}\ \bibnamefont {Linke}}, \bibinfo {author}
  {\bibfnamefont {T.~P.}\ \bibnamefont {Harty}}, \bibinfo {author}
  {\bibfnamefont {D.~P.~L.}\ \bibnamefont {Aude~Craik}}, \bibinfo {author}
  {\bibfnamefont {D.~N.}\ \bibnamefont {Stacey}}, \bibinfo {author}
  {\bibfnamefont {A.~M.}\ \bibnamefont {Steane}},\ and\ \bibinfo {author}
  {\bibfnamefont {D.~M.}\ \bibnamefont {Lucas}},\ }\href
  {https://doi.org/10.1038/nature16184} {\bibfield  {journal} {\bibinfo
  {journal} {Nature}\ }\textbf {\bibinfo {volume} {528}},\ \bibinfo {pages}
  {384} (\bibinfo {year} {2015})}\BibitemShut {NoStop}%
\bibitem [{\citenamefont {Malekakhlagh}\ \emph {et~al.}(2016)\citenamefont
  {Malekakhlagh}, \citenamefont {Petrescu},\ and\ \citenamefont
  {T{\"u}reci}}]{malekakhlagh2016}%
  \BibitemOpen
  \bibfield  {author} {\bibinfo {author} {\bibfnamefont {M.}~\bibnamefont
  {Malekakhlagh}}, \bibinfo {author} {\bibfnamefont {A.}~\bibnamefont
  {Petrescu}},\ and\ \bibinfo {author} {\bibfnamefont {H.~E.}\ \bibnamefont
  {T{\"u}reci}},\ }\href {https://doi.org/10.1103/PhysRevA.94.063848}
  {\bibfield  {journal} {\bibinfo  {journal} {Physical Review A}\ }\textbf
  {\bibinfo {volume} {94}},\ \bibinfo {pages} {063848} (\bibinfo {year}
  {2016})}\BibitemShut {NoStop}%
\bibitem [{\citenamefont {Sampaio}\ \emph {et~al.}(2017)\citenamefont
  {Sampaio}, \citenamefont {Suomela}, \citenamefont {Schmidt},\ and\
  \citenamefont {{Ala-Nissila}}}]{sampaio2017}%
  \BibitemOpen
  \bibfield  {author} {\bibinfo {author} {\bibfnamefont {R.}~\bibnamefont
  {Sampaio}}, \bibinfo {author} {\bibfnamefont {S.}~\bibnamefont {Suomela}},
  \bibinfo {author} {\bibfnamefont {R.}~\bibnamefont {Schmidt}},\ and\ \bibinfo
  {author} {\bibfnamefont {T.}~\bibnamefont {{Ala-Nissila}}},\ }\href
  {https://doi.org/10.1103/PhysRevA.95.022120} {\bibfield  {journal} {\bibinfo
  {journal} {Physical Review A}\ }\textbf {\bibinfo {volume} {95}},\ \bibinfo
  {pages} {022120} (\bibinfo {year} {2017})}\BibitemShut {NoStop}%
\bibitem [{\citenamefont {Bu{\v z}ek}\ \emph {et~al.}(1999)\citenamefont {Bu{\v
  z}ek}, \citenamefont {Drobn{\'y}}, \citenamefont {Kim}, \citenamefont
  {Havukainen},\ and\ \citenamefont {Knight}}]{buzek1999}%
  \BibitemOpen
  \bibfield  {author} {\bibinfo {author} {\bibfnamefont {V.}~\bibnamefont
  {Bu{\v z}ek}}, \bibinfo {author} {\bibfnamefont {G.}~\bibnamefont
  {Drobn{\'y}}}, \bibinfo {author} {\bibfnamefont {M.~G.}\ \bibnamefont {Kim}},
  \bibinfo {author} {\bibfnamefont {M.}~\bibnamefont {Havukainen}},\ and\
  \bibinfo {author} {\bibfnamefont {P.~L.}\ \bibnamefont {Knight}},\ }\href
  {https://doi.org/10.1103/PhysRevA.60.582} {\bibfield  {journal} {\bibinfo
  {journal} {Physical Review A}\ }\textbf {\bibinfo {volume} {60}},\ \bibinfo
  {pages} {582} (\bibinfo {year} {1999})}\BibitemShut {NoStop}%
\bibitem [{\citenamefont {Hansen}\ and\ \citenamefont
  {Tywoniuk}(2023)}]{hansen2023}%
  \BibitemOpen
  \bibfield  {author} {\bibinfo {author} {\bibfnamefont {J.~P.}\ \bibnamefont
  {Hansen}}\ and\ \bibinfo {author} {\bibfnamefont {K.}~\bibnamefont
  {Tywoniuk}},\ }\href {https://doi.org/10.1103/PhysRevA.108.053707} {\bibfield
   {journal} {\bibinfo  {journal} {Physical Review A}\ }\textbf {\bibinfo
  {volume} {108}},\ \bibinfo {pages} {053707} (\bibinfo {year}
  {2023})}\BibitemShut {NoStop}%
\bibitem [{\citenamefont {Lorenzo}\ \emph {et~al.}(2013)\citenamefont
  {Lorenzo}, \citenamefont {Plastina},\ and\ \citenamefont
  {Paternostro}}]{lorenzo2013}%
  \BibitemOpen
  \bibfield  {author} {\bibinfo {author} {\bibfnamefont {S.}~\bibnamefont
  {Lorenzo}}, \bibinfo {author} {\bibfnamefont {F.}~\bibnamefont {Plastina}},\
  and\ \bibinfo {author} {\bibfnamefont {M.}~\bibnamefont {Paternostro}},\
  }\href {https://doi.org/10.1103/PhysRevA.88.020102} {\bibfield  {journal}
  {\bibinfo  {journal} {Physical Review A}\ }\textbf {\bibinfo {volume} {88}},\
  \bibinfo {pages} {020102} (\bibinfo {year} {2013})}\BibitemShut {NoStop}%
\bibitem [{\citenamefont {Breuer}\ \emph {et~al.}(2009)\citenamefont {Breuer},
  \citenamefont {Laine},\ and\ \citenamefont {Piilo}}]{breuer2009}%
  \BibitemOpen
  \bibfield  {author} {\bibinfo {author} {\bibfnamefont {H.-P.}\ \bibnamefont
  {Breuer}}, \bibinfo {author} {\bibfnamefont {E.-M.}\ \bibnamefont {Laine}},\
  and\ \bibinfo {author} {\bibfnamefont {J.}~\bibnamefont {Piilo}},\ }\href
  {https://doi.org/10.1103/PhysRevLett.103.210401} {\bibfield  {journal}
  {\bibinfo  {journal} {Physical Review Letters}\ }\textbf {\bibinfo {volume}
  {103}},\ \bibinfo {pages} {210401} (\bibinfo {year} {2009})}\BibitemShut
  {NoStop}%
\bibitem [{\citenamefont {Kaufman}\ \emph {et~al.}(2015)\citenamefont
  {Kaufman}, \citenamefont {Lester}, \citenamefont {{Foss-Feig}}, \citenamefont
  {Wall}, \citenamefont {Rey},\ and\ \citenamefont {Regal}}]{kaufman2015}%
  \BibitemOpen
  \bibfield  {author} {\bibinfo {author} {\bibfnamefont {A.~M.}\ \bibnamefont
  {Kaufman}}, \bibinfo {author} {\bibfnamefont {B.~J.}\ \bibnamefont {Lester}},
  \bibinfo {author} {\bibfnamefont {M.}~\bibnamefont {{Foss-Feig}}}, \bibinfo
  {author} {\bibfnamefont {M.~L.}\ \bibnamefont {Wall}}, \bibinfo {author}
  {\bibfnamefont {A.~M.}\ \bibnamefont {Rey}},\ and\ \bibinfo {author}
  {\bibfnamefont {C.~A.}\ \bibnamefont {Regal}},\ }\href
  {https://doi.org/10.1038/nature16073} {\bibfield  {journal} {\bibinfo
  {journal} {Nature}\ }\textbf {\bibinfo {volume} {527}},\ \bibinfo {pages}
  {208} (\bibinfo {year} {2015})}\BibitemShut {NoStop}%
\bibitem [{\citenamefont {Shen}\ and\ \citenamefont {Fan}(2007)}]{shen2007}%
  \BibitemOpen
  \bibfield  {author} {\bibinfo {author} {\bibfnamefont {J.-T.}\ \bibnamefont
  {Shen}}\ and\ \bibinfo {author} {\bibfnamefont {S.}~\bibnamefont {Fan}},\
  }\href {https://doi.org/10.1103/PhysRevLett.98.153003} {\bibfield  {journal}
  {\bibinfo  {journal} {Physical Review Letters}\ }\textbf {\bibinfo {volume}
  {98}},\ \bibinfo {pages} {153003} (\bibinfo {year} {2007})}\BibitemShut
  {NoStop}%
\bibitem [{\citenamefont {Zhou}\ \emph {et~al.}(2008)\citenamefont {Zhou},
  \citenamefont {Gong}, \citenamefont {Liu}, \citenamefont {Sun},\ and\
  \citenamefont {Nori}}]{zhou2008}%
  \BibitemOpen
  \bibfield  {author} {\bibinfo {author} {\bibfnamefont {L.}~\bibnamefont
  {Zhou}}, \bibinfo {author} {\bibfnamefont {Z.~R.}\ \bibnamefont {Gong}},
  \bibinfo {author} {\bibfnamefont {Y.-x.}\ \bibnamefont {Liu}}, \bibinfo
  {author} {\bibfnamefont {C.~P.}\ \bibnamefont {Sun}},\ and\ \bibinfo {author}
  {\bibfnamefont {F.}~\bibnamefont {Nori}},\ }\href
  {https://doi.org/10.1103/PhysRevLett.101.100501} {\bibfield  {journal}
  {\bibinfo  {journal} {Physical Review Letters}\ }\textbf {\bibinfo {volume}
  {101}},\ \bibinfo {pages} {100501} (\bibinfo {year} {2008})}\BibitemShut
  {NoStop}%
\bibitem [{\citenamefont {Lehmberg}(1970)}]{lehmberg1970}%
  \BibitemOpen
  \bibfield  {author} {\bibinfo {author} {\bibfnamefont {R.~H.}\ \bibnamefont
  {Lehmberg}},\ }\href {https://doi.org/10.1103/PhysRevA.2.883} {\bibfield
  {journal} {\bibinfo  {journal} {Physical Review A}\ }\textbf {\bibinfo
  {volume} {2}},\ \bibinfo {pages} {883} (\bibinfo {year} {1970})}\BibitemShut
  {NoStop}%
\bibitem [{\citenamefont {Scully}\ and\ \citenamefont
  {Zubairy}(1997)}]{scully1997}%
  \BibitemOpen
  \bibfield  {author} {\bibinfo {author} {\bibfnamefont {M.~O.}\ \bibnamefont
  {Scully}}\ and\ \bibinfo {author} {\bibfnamefont {M.~S.}\ \bibnamefont
  {Zubairy}},\ }\href {https://doi.org/10.1017/CBO9780511813993} {\emph
  {\bibinfo {title} {Quantum {{Optics}}}}},\ \bibinfo {edition} {1st}\ ed.\
  (\bibinfo  {publisher} {Cambridge University Press},\ \bibinfo {year}
  {1997})\BibitemShut {NoStop}%
\bibitem [{\citenamefont {Ruggenthaler}\ \emph {et~al.}(2014)\citenamefont
  {Ruggenthaler}, \citenamefont {Flick}, \citenamefont {Pellegrini},
  \citenamefont {Appel}, \citenamefont {Tokatly},\ and\ \citenamefont
  {Rubio}}]{ruggenthaler2014}%
  \BibitemOpen
  \bibfield  {author} {\bibinfo {author} {\bibfnamefont {M.}~\bibnamefont
  {Ruggenthaler}}, \bibinfo {author} {\bibfnamefont {J.}~\bibnamefont {Flick}},
  \bibinfo {author} {\bibfnamefont {C.}~\bibnamefont {Pellegrini}}, \bibinfo
  {author} {\bibfnamefont {H.}~\bibnamefont {Appel}}, \bibinfo {author}
  {\bibfnamefont {I.~V.}\ \bibnamefont {Tokatly}},\ and\ \bibinfo {author}
  {\bibfnamefont {A.}~\bibnamefont {Rubio}},\ }\href
  {https://doi.org/10.1103/PhysRevA.90.012508} {\bibfield  {journal} {\bibinfo
  {journal} {Physical Review A}\ }\textbf {\bibinfo {volume} {90}},\ \bibinfo
  {pages} {012508} (\bibinfo {year} {2014})}\BibitemShut {NoStop}%
\bibitem [{\citenamefont {Baba}(2008)}]{baba2008}%
  \BibitemOpen
  \bibfield  {author} {\bibinfo {author} {\bibfnamefont {T.}~\bibnamefont
  {Baba}},\ }\href {https://doi.org/10.1038/nphoton.2008.146} {\bibfield
  {journal} {\bibinfo  {journal} {Nature Photonics}\ }\textbf {\bibinfo
  {volume} {2}},\ \bibinfo {pages} {465} (\bibinfo {year} {2008})}\BibitemShut
  {NoStop}%
\bibitem [{\citenamefont {Horodecki}\ \emph {et~al.}(1995)\citenamefont
  {Horodecki}, \citenamefont {Horodecki},\ and\ \citenamefont
  {Horodecki}}]{horodecki1995}%
  \BibitemOpen
  \bibfield  {author} {\bibinfo {author} {\bibfnamefont {R.}~\bibnamefont
  {Horodecki}}, \bibinfo {author} {\bibfnamefont {P.}~\bibnamefont
  {Horodecki}},\ and\ \bibinfo {author} {\bibfnamefont {M.}~\bibnamefont
  {Horodecki}},\ }\href {https://doi.org/10.1016/0375-9601(95)00214-N}
  {\bibfield  {journal} {\bibinfo  {journal} {Physics Letters A}\ }\textbf
  {\bibinfo {volume} {200}},\ \bibinfo {pages} {340} (\bibinfo {year}
  {1995})}\BibitemShut {NoStop}%
\bibitem [{\citenamefont {Simon}(2000)}]{simon2000}%
  \BibitemOpen
  \bibfield  {author} {\bibinfo {author} {\bibfnamefont {R.}~\bibnamefont
  {Simon}},\ }\href {https://doi.org/10.1103/PhysRevLett.84.2726} {\bibfield
  {journal} {\bibinfo  {journal} {Physical Review Letters}\ }\textbf {\bibinfo
  {volume} {84}},\ \bibinfo {pages} {2726} (\bibinfo {year}
  {2000})}\BibitemShut {NoStop}%
\bibitem [{\citenamefont {Mazzola}\ \emph {et~al.}(2010)\citenamefont
  {Mazzola}, \citenamefont {Piilo},\ and\ \citenamefont
  {Maniscalco}}]{mazzola2010}%
  \BibitemOpen
  \bibfield  {author} {\bibinfo {author} {\bibfnamefont {L.}~\bibnamefont
  {Mazzola}}, \bibinfo {author} {\bibfnamefont {J.}~\bibnamefont {Piilo}},\
  and\ \bibinfo {author} {\bibfnamefont {S.}~\bibnamefont {Maniscalco}},\
  }\href {https://doi.org/10.1103/PhysRevLett.104.200401} {\bibfield  {journal}
  {\bibinfo  {journal} {Physical Review Letters}\ }\textbf {\bibinfo {volume}
  {104}},\ \bibinfo {pages} {200401} (\bibinfo {year} {2010})}\BibitemShut
  {NoStop}%
\bibitem [{\citenamefont {Walter}\ \emph {et~al.}(2017)\citenamefont {Walter},
  \citenamefont {Kurpiers}, \citenamefont {Gasparinetti}, \citenamefont
  {Magnard}, \citenamefont {Potocnik}, \citenamefont {Salath{\'e}},
  \citenamefont {Pechal}, \citenamefont {Mondal}, \citenamefont {Oppliger},
  \citenamefont {Eichler},\ and\ \citenamefont {Wallraff}}]{walter2017}%
  \BibitemOpen
  \bibfield  {author} {\bibinfo {author} {\bibfnamefont {T.}~\bibnamefont
  {Walter}}, \bibinfo {author} {\bibfnamefont {P.}~\bibnamefont {Kurpiers}},
  \bibinfo {author} {\bibfnamefont {S.}~\bibnamefont {Gasparinetti}}, \bibinfo
  {author} {\bibfnamefont {P.}~\bibnamefont {Magnard}}, \bibinfo {author}
  {\bibfnamefont {A.}~\bibnamefont {Potocnik}}, \bibinfo {author}
  {\bibfnamefont {Y.}~\bibnamefont {Salath{\'e}}}, \bibinfo {author}
  {\bibfnamefont {M.}~\bibnamefont {Pechal}}, \bibinfo {author} {\bibfnamefont
  {M.}~\bibnamefont {Mondal}}, \bibinfo {author} {\bibfnamefont
  {M.}~\bibnamefont {Oppliger}}, \bibinfo {author} {\bibfnamefont
  {C.}~\bibnamefont {Eichler}},\ and\ \bibinfo {author} {\bibfnamefont
  {A.}~\bibnamefont {Wallraff}},\ }\href
  {https://doi.org/10.1103/PhysRevApplied.7.054020} {\bibfield  {journal}
  {\bibinfo  {journal} {Physical Review Applied}\ }\textbf {\bibinfo {volume}
  {7}},\ \bibinfo {pages} {054020} (\bibinfo {year} {2017})}\BibitemShut
  {NoStop}%
\bibitem [{\citenamefont {Gualtieri}\ \emph {et~al.}(1994)\citenamefont
  {Gualtieri}, \citenamefont {Kosinski},\ and\ \citenamefont
  {Ballato}}]{gualtieri1994}%
  \BibitemOpen
  \bibfield  {author} {\bibinfo {author} {\bibfnamefont {J.}~\bibnamefont
  {Gualtieri}}, \bibinfo {author} {\bibfnamefont {J.}~\bibnamefont
  {Kosinski}},\ and\ \bibinfo {author} {\bibfnamefont {A.}~\bibnamefont
  {Ballato}},\ }\href {https://doi.org/10.1109/58.265820} {\bibfield  {journal}
  {\bibinfo  {journal} {IEEE Transactions on Ultrasonics, Ferroelectrics, and
  Frequency Control}\ }\textbf {\bibinfo {volume} {41}},\ \bibinfo {pages} {53}
  (\bibinfo {year} {1994})}\BibitemShut {NoStop}%
\bibitem [{\citenamefont {Masmanidis}\ \emph {et~al.}(2007)\citenamefont
  {Masmanidis}, \citenamefont {Karabalin}, \citenamefont {De~Vlaminck},
  \citenamefont {Borghs}, \citenamefont {Freeman},\ and\ \citenamefont
  {Roukes}}]{masmanidis2007}%
  \BibitemOpen
  \bibfield  {author} {\bibinfo {author} {\bibfnamefont {S.~C.}\ \bibnamefont
  {Masmanidis}}, \bibinfo {author} {\bibfnamefont {R.~B.}\ \bibnamefont
  {Karabalin}}, \bibinfo {author} {\bibfnamefont {I.}~\bibnamefont
  {De~Vlaminck}}, \bibinfo {author} {\bibfnamefont {G.}~\bibnamefont {Borghs}},
  \bibinfo {author} {\bibfnamefont {M.~R.}\ \bibnamefont {Freeman}},\ and\
  \bibinfo {author} {\bibfnamefont {M.~L.}\ \bibnamefont {Roukes}},\ }\href
  {https://doi.org/10.1126/science.1144793} {\bibfield  {journal} {\bibinfo
  {journal} {Science}\ }\textbf {\bibinfo {volume} {317}},\ \bibinfo {pages}
  {780} (\bibinfo {year} {2007})}\BibitemShut {NoStop}%
\bibitem [{\citenamefont {Thorbeck}\ \emph {et~al.}(2023)\citenamefont
  {Thorbeck}, \citenamefont {Eddins}, \citenamefont {Lauer}, \citenamefont
  {McClure},\ and\ \citenamefont {Carroll}}]{thorbeck2023}%
  \BibitemOpen
  \bibfield  {author} {\bibinfo {author} {\bibfnamefont {T.}~\bibnamefont
  {Thorbeck}}, \bibinfo {author} {\bibfnamefont {A.}~\bibnamefont {Eddins}},
  \bibinfo {author} {\bibfnamefont {I.}~\bibnamefont {Lauer}}, \bibinfo
  {author} {\bibfnamefont {D.~T.}\ \bibnamefont {McClure}},\ and\ \bibinfo
  {author} {\bibfnamefont {M.}~\bibnamefont {Carroll}},\ }\href
  {https://doi.org/10.1103/PRXQuantum.4.020356} {\bibfield  {journal} {\bibinfo
   {journal} {PRX Quantum}\ }\textbf {\bibinfo {volume} {4}},\ \bibinfo {pages}
  {020356} (\bibinfo {year} {2023})}\BibitemShut {NoStop}%
\end{thebibliography}
